\UseRawInputEncoding
\documentclass[journal=jacsat,keywords=true,manuscript=article]{achemso}
\setkeys{acs}{etalmode=truncate,maxauthors=0,usetitle=true}

\usepackage{amsmath}
\usepackage{amssymb}
\usepackage{graphicx}
\usepackage{overpic}

\usepackage{xr-hyper}
\usepackage{hyperref}

\usepackage{geometry}
\usepackage{float} 
\usepackage{subfigure}
\usepackage[version=4]{mhchem}
\usepackage{textcomp}

\title{Differentiable OPLS Force Field Parameterization for Ionic Electrolytes and High-Throughput Application to Lithium-ion Batteries}
\author{Haichao Huang}
\affiliation{Tsinghua Shenzhen International Graduate School, Tsinghua University, Shenzhen, 518055 China}
\altaffiliation{These authors have contributed equally}

\author{Zilin Chen}
\affiliation{Tsinghua Shenzhen International Graduate School, Tsinghua University, Shenzhen, 518055 China}
\altaffiliation{These authors have contributed equally}

\author{Qi Liu}
\affiliation{Tsinghua Shenzhen International Graduate School, Tsinghua University, Shenzhen, 518055 China}
\altaffiliation{These authors have contributed equally}

\author{Tianqi Zhao}
\affiliation{Shanghai Smart Logic Technology, Shanghai 200443 China}

\author{Yunpei Liu}
\affiliation{Shanghai Smart Logic Technology, Shanghai 200443 China}

\author{Guotao Qiu}
\affiliation{CATL 21C Innovation Laboratory, Contemporary Amperex Technology Ltd., Ningde, 352100 China}

\author{Jianhui Chen}
\affiliation{Department of Physics, Institute of Quantum Science and Technology, Shanghai University, Shanghai, 200444, China}
\alsoaffiliation{CATL 21C Innovation Laboratory, Contemporary Amperex Technology Ltd., Ningde, 352100 China}

\author{Zhen Li}
\affiliation{Shanghai Smart Logic Technology, Shanghai 200443 China}

\author{Wenshuo Liang}
\affiliation{Shanghai Smart Logic Technology, Shanghai 200443 China}

\author{Minsung Cho}
\affiliation{Tsinghua Shenzhen International Graduate School, Tsinghua University, Shenzhen, 518055 China}

\author{Manxue Zhang}
\affiliation{Tsinghua Shenzhen International Graduate School, Tsinghua University, Shenzhen, 518055 China}

\author{Feiyu Kang}
\affiliation{Tsinghua Shenzhen International Graduate School, Tsinghua University, Shenzhen, 518055 China}

\author{Xiaolong Zou}
\affiliation{Tsinghua Shenzhen International Graduate School, Tsinghua University, Shenzhen, 518055 China}

\author{Yidan Cao}
\affiliation{Shenzhen All-Solid-State Lithium Battery Electrolyte Engineering Research Center, Institute of Materials Research (IMR), Tsinghua Shenzhen International Graduate School, Tsinghua University, Shenzhen, 518055 China}
\email{yidancao@sz.tsinghua.edu.cn}

\author{Xushan Zhao}
\affiliation{CATL 21C Innovation Laboratory, Contemporary Amperex Technology Ltd., Ningde, 352100 China}
\email{ZhaoXS@catl-21c.com}

\author{Ziqi Cheng}
\affiliation{Shanghai Smart Logic Technology, Shanghai 200443 China}
\email{ziqi.cheng@smartlogictech.com}

\author{Ye Mei}
\affiliation{Shanghai Smart Logic Technology, Shanghai 200443 China}

\begin{document}

\maketitle



\newpage

\begin{abstract}

The rational design of ionic electrolytes in lithium-ion batteries (LIBs) is severely hampered by the vast combinatorial space of solvents and salts, and the low efficiency of empirical trial-and-error approaches. Molecular dynamics (MD) serves as a critical bridge between microscopic solvation structures and macroscopic physicochemical properties, but classical force fields often lack accuracy for multicomponent electrolytes. 
To address these challenges, an automated differentiable parameterization workflow for the OPLS-AA force field, specifically tailored to general ionic electrolytes, was developed. This workflow adopts topology-guided atom typification to reduce parameter redundancy and optimizes Lennard-Jones parameters via the DMFF framework, in which experimentally measured densities are used as the fitting target and ionic conductivity as an independent validation metric. Through rigorous convergence tests, a standardized simulation protocol featuring $\sim$100,000 atom systems and 35-40 ns NVT runs was established to ensure reliable quantification of the transport properties of electrolytes.
Furthermore, high-throughput MD simulations were performed on the Tianqiong platform for over 10,000 electrolyte formulations that encompass 67 solvents and 15 lithium salts, generating a comprehensive dataset for five key electrolyte properties consisting of density, dielectric constant, viscosity, diffusion coefficient, and ionic conductivity. The t-SNE visualization reveals partial clustering of distinct salt chemistries, continuous property gradients as functions of salt concentration and temperature, and internal physical self-consistency across the measured electrolyte parameters—while simultaneously identifying solvent components as another critical performance determinant. Collectively, the accurate and transferable force field and the rich data resource establish a robust foundation for future data-driven design of ionic electrolytes.
\end{abstract}

\section{Introduction}

Among various energy storage devices, LIBs have dominated the portable electronics and electric vehicle markets since 1991\cite{bedrov_molecular_2019}, owing to their outstanding energy density\cite{Liang_review_2019}, long cycle life\cite{Li_single_2023,weber_long_2019}, and low self-discharge rate\cite{chaudhari_challenges_2025}.  While electrode materials set the theoretical upper limit of energy density, the actual rate performance, accessible capacity, and long-term durability are largely determined by the electrolyte properties and the electrode-electrolyte interfacial stability\cite{yao_applying_2022}. As a key element of LIBs, the electrolyte acts as both an ionic conductor and an electronic insulator\cite{yao_applying_2022}, which dictates the rate capability, cycling stability, and safety of practical batteries\cite{kumar_electrolytomics_2025}. Commercial liquid electrolytes are composed of polar solvents, lithium salts, and functional additives. For commonly used solvents, cyclic carbonates often show strong \ce{Li+} solvation capability, high conductivity, and high viscosity, while linear carbonates, esters, and ethers exhibit low viscosity and moderate solvation ability and conductivity\cite{borodin_competitive_2016}. After three decades of exploration, extensive experiments have been carried out to measure key properties of electrolytes, including ionic conductivity and Coulombic efficiency. However, efficient screening paradigms are still lacking, and public databases remain sparse—offering only approximately 10,000 reliable conductivity data points that cover a minuscule fraction of feasible electrolyte compositions\cite{kumar_electrolytomics_2025}. This scarcity severely constrains data-driven materials discovery and impedes the development of high-performance electrolytes.\cite{yao_applying_2022,kumar_electrolytomics_2025,yang_unified_2026}.

In contrast to the inefficiency and expense of experimental measurements, MD simulations deliver quantitative predictions of diverse physicochemical properties with outstanding reproducibility and consistency, rendering them better suited for high-throughput screening. In addition, with the powerful ability to bridge microscopic structures and macroscopic properties, MD simulations can be used to reveal the atomic-level mechanisms of ionic solvation, transport, and interfacial behaviors of electrolytes\cite{bedrov_molecular_2019}. MD simulations are categorized into classical MD (CMD), \textit{ab initio} MD (AIMD), and machine learning MD (MLMD) according to different approaches to the interatomic interactions\cite{yao_applying_2022}. AIMD offers high accuracy at prohibitive computational cost, limiting the system size and simulation time to unrealistic scales. MLMD balances accuracy and efficiency but suffers from poor transferability outside training domains, making it challenging for large-scale electrolyte screening. In comparison, classical force fields such as OPLS\cite{jorgensen_development_1996}, CHARMM\cite{vanommeslaeghe_charmm_2010,cornell_second_1995}, and AMBER\cite{mackerell_all-atom_1998} enable fast simulations of large systems with millions of atoms, retaining irreplaceable advantages in high-throughput computational screening.

The OPLS force field series is widely used in electrolyte simulations owing to its reliable description of organic liquids and comprehensive coverage of atom types\cite{doherty_revisiting_2017}. However, ready-to-use generic force fields often fail to achieve satisfactory accuracy for multicomponent electrolytes, especially for transport properties including viscosity, diffusion coefficient, and ionic conductivity\cite{ma_prediction_2017,liu_thermal_2012,kowsari_molecular_2009,rey-castro_transport_2006,sprenger_general_2015}. Direct application of generic parameters yields substantial deviations from experimental measurements, underscoring the need for targeted parameter optimization. Unfortunately, force field parameterization for commercial electrolytes is highly challenging, stemming from the diverse chemical environments of solvent and salt components, the large pool of tunable parameters, and the absence of automated optimization workflows. Recent development of the Differentiable Molecular Force Field (DMFF)\cite{wang_dmff_2023} framework provides a transformative strategy for automated force field optimization. By constructing end-to-end differentiable simulation pipelines and leveraging trajectory reweighting techniques, DMFF enables gradient-based parameter refinement using direct experimental observables, greatly improving optimization efficiency and accuracy. This framework allows systematic tuning of force field parameters to match experimental properties without manual intervention, offering an ideal solution to address the shortcomings of conventional classical force fields for electrolyte simulations.

Even with an accurate and efficient force field, high-throughput electrolyte screening remains constrained by a critical computational bottleneck. As shown in our convergence analysis of system size, accurate prediction of transport properties (e.g., conductivity, viscosity) requires sufficiently large simulation systems ($\ge 10^5$ atoms) to mitigate the finite-size effects and reduce statistical uncertainty. However, the computational cost of large-scale MD simulations grows superlinearly with increasing system size\cite{allen_statistical_2017,darden_particle_1993}. While conventional computing clusters are capable of running fast MD calculations, they lack specialized optimization for the fine-grained parallelism and long-range electrostatic calculations intrinsic to MD simulations, resulting in low efficiency and high resource consumption for high-throughput screening. To address this challenge, the Tianqiong platform\cite{Tianqiong} was employed in this work, which is specifically designed for molecular dynamics simulations and enables efficient high-throughput simulations of electrolyte systems with more than $10^5$ atoms.

In this work, an automated differentiable optimization workflow for the OPLS-AA force field\cite{jorgensen_development_1996,sambasivarao_development_2009} has been developed, specifically tailored to general ionic liquid electrolytes. Topology-guided atom typification on high-occurrence solvent and salt molecules from the EDB-1 database\cite{kumar_electrolytomics_2025} has been performed to reduce parameter redundancy and improve transferability. The Lennard-Jones (LJ) parameters\cite{schwerdtfeger_100_2024} are optimized via DMFF\cite{wang_dmff_2023}, with experimental density as the fitting target and ionic conductivity as an independent validation metric. Finite-size effects and convergence behavior of transport properties have been rigorously investigated to ensure reliable prediction of viscosity, diffusion coefficient, and ionic conductivity. High-throughput simulations of over 10,000 electrolyte formulations have been conducted on the Tianqiong platform, and a comprehensive property dataset with broad compositional coverage and high data reliability has been generated therefrom. In addition to a versatile, accurate, and transferable OPLS force field and a reliable LIBs electrolyte dataset\cite{borodin_competitive_2016}, a scalable paradigm for data-driven force field development and high-throughput electrolyte screening has been established in this work.

\section{Models and Computational Methods}

\subsection{Overall Workflow}
\begin{figure}[H]
\centering
  \includegraphics[width=0.9\textwidth]{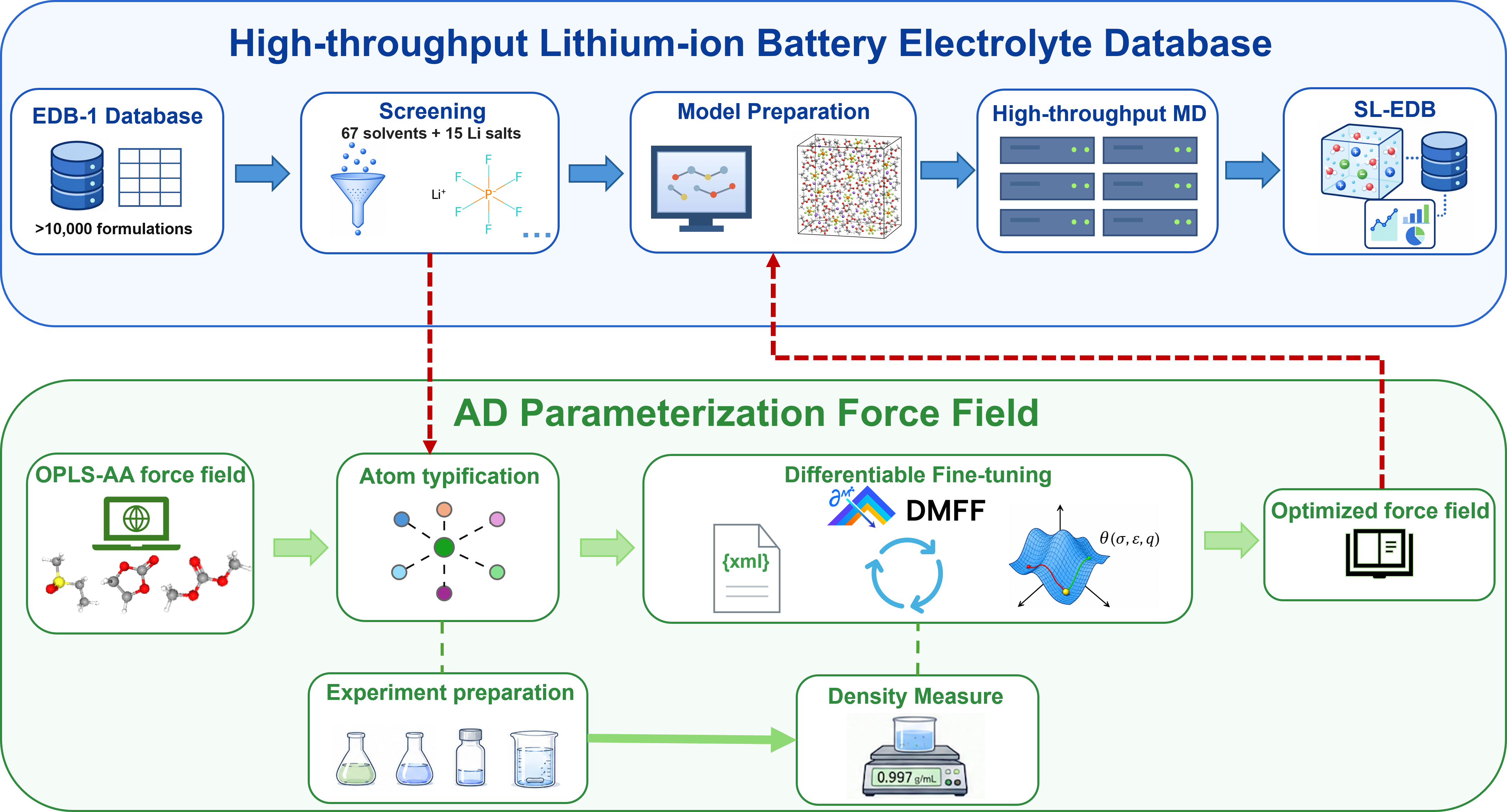}
  \caption{Workflow for transferable electrolyte force field optimization and high-throughput physicochemical property dataset construction.} 
  \label{fig:workflow}
\end{figure}

This work establishes an integrated closed-loop workflow that couples transferable force field optimization with high-throughput electrolyte property dataset construction, as schematically illustrated in Figure~\ref{fig:workflow}. The workflow comprises two interconnected modules: differentiable optimization of a general OPLS-based electrolyte force field, and large-scale MD-driven dataset construction. The component space defined by the dataset delineates the scope of atom types covered for force field parameterization, while the optimized force field, in turn, provides a reliable computational foundation for high-throughput property predictions, thereby forming a mutually reinforcing closed loop.

In the force field optimization module, the classical OPLS-AA all-atom force field serves as the initial parameter framework. To reduce redundancy and enhance transferability across diverse electrolyte systems, all atom types of the selected solvent and salt components are reclassified according to their local chemical environments, including coordination number, neighboring element types, and hybridization states. This reclassification condenses the original redundant atom types into 68 distinct types, with 136 Lennard-Jones (LJ) parameters subject to subsequent optimization. A representative formulation set encompassing all classified atom types was then constructed, and automated parameter optimization was performed via the Differentiable Molecular Force Field (DMFF) framework. Using experimental density as the fitting target, DMFF enables efficient gradient-based updates of LJ parameters through trajectory reweighting, thereby circumventing the need for repeated, computationally expensive MD simulations during tuning. Finally, ionic conductivity is utilized as an independent validation metric to ensure that the optimized force field achieves balanced accuracy in both structural and transport properties, thus avoiding overfitting solely to density. The resulting universal force field demonstrates excellent transferability across most mainstream commercial electrolyte systems.

For the electrolyte property dataset module, the public EDB-1 liquid electrolyte conductivity dataset\cite{kumar_electrolytomics_2025} serves as the foundation for component screening. Frequency-based analysis of solvents and lithium salts in the EDB-1 database identifies a core set of 67 solvents and 15 salts, spanning mainstream carbonate-based systems and commonly used functional electrolyte additives. From this selection, a series of simulation systems is constructed, encompassing both single-salt and dual-salt formulations with systematically varied solvent ratios and salt concentrations.

Prior to large-scale calculations, simulation parameters are validated via convergence checks of representative physicochemical properties, confirming that a system size of $10^{5}$ atoms and a simulation duration of 35-40 ns suffice for reliable transport-property evaluation. Leveraging the exceptional computational efficiency of the Tianqiong platform for large-scale MD simulations, high-throughput calculations were conducted over more than 10,000 electrolyte formulations. For each formulation, five main physicochemical properties are computed: density, ionic conductivity, \ce{Li+} self-diffusion coefficient, shear viscosity, and relative dielectric constant, ultimately yielding a comprehensive electrolyte property dataset with broad component coverage and robust data reliability.

This integrated automation workflow substantially reduces the manual effort of conventional force-field tuning and batch simulations, while simultaneously providing a robust methodological and data foundation for subsequent data-driven formulation design and performance optimization.

\subsection{General Force Field Fine-tuning}
General all-atom force fields such as OPLS-AA provide broad molecular coverage. However, they exhibit systematic deviations in describing strong ionic-solvent interactions in concentrated lithium battery electrolytes, leading to inaccurate predictions of ion association and transport. To address this limitation, a refined general force field for multicomponent electrolytes is developed on the basis of OPLS-AA through three steps: atom typification, theoretical model construction, and differentiable gradient-based optimization. This subsection details the corresponding fine-tuning workflow.

\subsubsection{Atom Typification}
The original OPLS-AA force field adopts an overly fine-grained atom-typing scheme, which leads to high parameter redundancy and inadequate data support for rare electrolyte-relevant atom types. To address this issue, atom types are reclassified according to local chemical environments, as defined by three criteria: (1) the number of first-nearest neighboring atoms, (2) the element types of these neighbors, and (3) the atomic hybridization state (e.g., $sp$/$sp^2$/$sp^3$). Atoms satisfying all three criteria are assigned to an identical atom type. This topology-guided reclassification effectively reduces redundancy in the original OPLS-AA typing and enhances parameter transferability across diverse electrolyte systems.

For atoms that are either experimentally inaccessible or exert negligible influence on bulk properties (e.g., buried \ce{Si} atoms), the original OPLS-AA parameters are retained. Chemically analogous atom types-such as $\alpha$-fluorinated carbons and monofluorinated secondary carbons-are manually mapped to equivalent alkane quaternary carbons to further reduce parameter count. Through this classification and merging strategy, the initial 78 atom types are condensed into 68 distinct types, yielding 136 Lennard-Jones parameters (68 $\sigma/\epsilon$ pairs).

\subsubsection{Force Field Model}
The rationality of the initial model directly governs both the convergence speed of optimization and the final generalization capability. A concentrated-electrolyte-compatible model was therefore developed on the basis of OPLS-AA, incorporating charge scaling\cite{cui_influence_2019,liang_charge_2024,fan_charge_2025,avula_efficient_2021} and a dual-criterion fitting strategy. OPLS-AA was chosen as the starting framework owing to its proven performance in predicting conformational energies and thermodynamic properties of organic molecules. Parameters for most solvents and salts were sourced from the LigParGen server\cite{dodda_114cm1a-lbcc_2017,dodda_ligpargen_2017,jorgensen_potential_2005} using the 
$1.14\times$CM1A charge model\cite{dodda_114cm1a-lbcc_2017,dodda_ligpargen_2017}. For components not supported by LigParGen (e.g., nitriles, borates, and anions), parameters were derived from high-quality literature, supplemented by MP2/cc-pVTZ quantum chemistry (QM) calculations or transfer from chemically analogous atoms\cite{storer_class_1995}.
Classical fixed-point charge models tend to overestimate Coulombic interactions in concentrated electrolytes, leading to excessive ion pairing and underestimated conductivity\cite{fan_charge_2025,avula_efficient_2021}. To mitigate this issue, adjustable charge scaling factors (0.7, 0.8, and 1.0) were applied to the charges of lithium salts. Although charge scaling has a marginal effect on density, it substantially improves conductivity predictions by effectively accounting for charge-screening effects. A factor of 0.7 was found to be optimal for most systems. The Lennard-Jones (LJ) parameters ($\sigma$ and $\epsilon$) were selected as the main optimization targets, as they dominate the short-range van der Waals interactions that determine the microstructure of the electrolyte. The experimental density-the most accurately measurable thermodynamic property-was adopted as the main fitting objective, while the ionic conductivity served as an independent validation metric. This dual-property validation approach prevents overfitting to density alone and ensures that the optimized force field accurately reproduces both structural and transport properties.

\subsubsection{Optimization}
Traditional force field optimization, which relies on trial-and-error or finite-difference gradients, is computationally expensive and poorly scalable. To overcome this limitation, the DMFF (Differentiable Molecular Force Field) program\cite{wang_dmff_2023} was used, combined with a grouped cyclic optimization strategy\cite{feng_screening_nodate}. DMFF enables automatic differentiation of the energy function via trajectory reweighting, allowing gradient evaluation of the loss function from precomputed MD trajectories. This avoids costly repeated simulations during optimization, substantially reducing computational overhead and enabling simultaneous optimization of hundreds of parameters.
To ensure broad applicability across all electrolyte systems, a group-by-group iterative protocol was implemented: (1) electrolyte formulations are divided into batches of eight; (2) MD simulations are performed for each batch using the current force field parameters; (3) the loss function-defined as the mean squared deviation between simulated and experimental density is computed, and LJ parameters are updated via gradient descent; (4) the optimized parameters serve as the initial guess for the subsequent batch; and (5) this cycle is repeated for at least 20 iterations until the total loss function converges (drift $<$ 5\%). This cyclic strategy ensures gradual convergence to an optimal parameter set applicable to all selected formulations, effectively balancing properties across systems while preventing overfitting to individual compositions and guaranteeing universal force field transferability.

\subsection{Electrolyte Property dataset Setting}

\subsubsection{Electrolyte Screening}
The EDB-1 electrolyte dataset was selected as the fundamental data source for formula screening, which covers diverse electrolyte systems including different solvent types (carbonates, ethers, esters), lithium salts (\ce{LiPF6}, \ce{LiFSI}, \ce{LiTFSI}, \ce{LiBOB}), salt concentrations, and temperatures.
Occurrence frequencies of each solvent and salt in EDB-1 are analyzed, and the high-frequency components are selected as the core constituents of the dataset. Specifically, 67 solvents and 15 salts with the highest occurrence counts are retained (see Figure S5 in Supporting Information), while low-frequency salts (e.g., \ce{Na}-based salts, imidazolium salts) and rare components with unobtainable initial force field parameters are excluded. This screening strategy allows the dataset to achieve comprehensive coverage of the main commercial electrolyte systems and align with industrial application scenarios\cite{borodin_competitive_2016}.
\subsubsection{Model Preparing}

In the modeling preparation stage, essential components for high-throughput computations are established, including experimental validation formulations, convergence tests, initial force field generation, and benchmark MD simulations on the Tianqiong platform.

\begin{itemize}
    \item \textbf{Formula Preparation} --- Based on the high-frequency components screened from EDB-1, typical single-salt binary-solvent formulations are extracted, and the formulation space is further expanded by varying solvent volume ratios (5:5 and 3:7) and salt concentrations (0.3 $\mathrm{M}$, 0.5 $\mathrm{M}$, and 1.0 $\mathrm{M}$). EC/DMC is used as the base solvent system, and other solvents are added as modifiers. The formulations cover all atom types involved in the force field optimization to ensure the universality of the optimized parameters.

    \item \textbf{Simulation Model Preparation} --- The system-size dependence and temporal convergence of transport properties (diffusion, conductivity, and viscosity) are rigorously assessed. A system of $10^5$ atoms and a simulation window of 30-40 ns are found to be necessary to eliminate finite-size artifacts and minimize statistical uncertainties, thereby ensuring robust predictions of transport properties.
    
    \item \textbf{Force Field Preparation} --- The initial OPLS-AA force field is used as the starting point, and DMFF is employed to fine-tune the force field parameters to match experimental density. The optimized force field is validated against experimental ionic conductivity from EDB-1 to ensure a balance between structural and transport-property predictions.
    
    \item \textbf{Workhorse Preparation} --- The prediction accuracy of traditional GPUs and the Tianqiong platform was compared for large-scale electrolyte MD simulations and found to be essentially equivalent. However, Tianqiong exhibited significantly superior computational efficiency for large-system simulations, and was therefore adopted as the primary platform for high-throughput calculations.
    \end{itemize}

\subsubsection{Property Prediction}
Large-scale MD simulations are performed on Tianqiong to compute five key electrolyte properties, with standardized calculation methods as follows. Density is a fundamental physical property of electrolytes, which directly determines the volumetric energy density of batteries and serves as a prerequisite for calculating other transport properties. In this work, densities are directly derived from the equilibrium trajectories of NPT ensemble simulations. All subsequent transport properties are calculated from NVT ensemble simulations starting from the equilibrated configurations obtained from NPT runs.
The self-diffusion coefficient $D$ of \ce{Li+} is a critical transport property that directly reflects lithium-ion mobility and governs battery rate capability. It is conventionally derived from the mean square displacement (MSD) via the Einstein relation\cite{allen_statistical_2017}:
\begin{equation}
D = \lim_{t \to \infty} \frac{1}{6tN} \sum_{i=1}^{N} \big\langle \big|\boldsymbol{r}_i(t) - \boldsymbol{r}_i(0)\big|^2 \big\rangle,
\end{equation}
in which $\boldsymbol{r}_i(t)$ and $\boldsymbol{r}_i(0)$ are the position vectors of the $i$-th particle at time $t$ and time zero, $N$ is the total number of particles of the target species, and $\langle \cdot \rangle$ is the ensemble average over equilibrium trajectories.

Ionic conductivity is a key physicochemical property of electrolytes that strongly influences the ohmic resistance and rate capability of battery systems. It is the primary target for electrolyte formulation optimization. Several methods are available for computing ionic conductivity in electrolytes. The Nernst-Einstein (NE) relation is a simple approach that calculates conductivity using the self-diffusion coefficients of charged ions\cite{gullbrekken_charge_2023,kowsari_molecular_2009}:
\begin{equation}
    \sigma_i^{\text{NE}} = \frac{z_i^2 e^2}{6 k_B T V} \lim_{t \to \infty} \frac{\mathrm{d}}{\mathrm{d}t} \left\langle \sum_{k=1}^{N_i} \left( \mathbf{r}_{k,i}(t) - \mathbf{r}_{k,i}(0) \right)^2 \right\rangle,
\end{equation}
in which $z_i$ is valence of species $i$, $e$ is elementary charge, $k_\mathrm{B}$ is the Boltzmann constant, $T$ is thermodynamic temperature, and $V$ is volume of the simulation box.
However, this method is only rigorous for ideal dilute solutions and neglects the diffusion correlations between positive and negative ions that are ubiquitous in real electrolytes, leading to overestimated conductivity values. Therefore, the Onsager transport theory\cite{wei_quantum_2025,gullbrekken_charge_2023} was adopted, which explicitly accounts for ionic correlations, to perform more accurate calculations. The partial ionic conductivity contribution from the correlation between species \(i\) and \(j\) is given by:
\begin{equation}
\sigma_{ij} = \frac{e^2}{6 k_\mathrm{B} T V}
\lim_{t \to \infty} \frac{\mathrm{d}}{\mathrm{d}t}
\left\langle
\sum_{k=1}^{N_i} \sum_{l=1}^{N_j} z_i z_j \left[ \mathbf{r}_{k,i}(t) - \mathbf{r}_{k,i}(0) \right]
\cdot
\left[ \mathbf{r}_{l,j}(t) - \mathbf{r}_{l,j}(0) \right]
\right\rangle.
\label{eq:lij}
\end{equation}
The total ionic conductivity is obtained by summing over all ionic pairs. For a binary electrolyte (one cation and one anion), the equation simplifies to:
\begin{equation}
\sigma = \sigma_{++} + \sigma_{--} + 2\sigma_{+-},
\end{equation}      
where $\sigma_{++}$, $\sigma_{--}$, and $\sigma_{+-}$ denote the partial ionic conductivity contributions from cation-cation correlations, anion-anion correlations, and cation-anion correlations, respectively. In the dilute limit where ionic correlations vanish, $\sigma_{+-}$ approaches zero, and $\sigma_{++}$, $\sigma_{--}$ converge to $\sigma_+^{\text{NE}}$ and $\sigma_-^{\text{NE}}$, respectively.

Shear viscosity is another key rheological property of electrolytes, which directly affects the migration rate of ions and the wettability of electrolytes in electrode pores, thereby influencing the rate performance and low-temperature performance of batteries. Viscosity was calculated using the Green-Kubo relation\cite{rey-castro_transport_2006,ma_prediction_2017,liu_thermal_2012,yao_probing_2023,yao_probing_2023,zhang_reliable_2015} based on the virial stress autocorrelation function:
\begin{equation}
\eta = \frac{V}{3 k_\mathrm{B} T}
\int_0^\infty \left[
\left\langle P_{xy}(0) P_{xy}(t) \right\rangle +
\left\langle P_{xz}(0) P_{xz}(t) \right\rangle +
\left\langle P_{yz}(0) P_{yz}(t) \right\rangle
\right] \mathrm{d}t,
\label{eq:viscosity}
\end{equation}
in which $P_{\alpha\beta}$ is the off-diagonal component of the virial stress tensor.
Relative dielectric constant is an important parameter reflecting the ability of solvent molecules to screen Coulomb interactions. It directly determines the dissociation degree of lithium salts, the structure of ion solvation shells, and the extent of ion association, making it an important basis for electrolyte formulation design. It was computed using the Einstein fluctuation formula\cite{yao_applying_2022,bedrov_molecular_2019}:
\begin{equation}
\varepsilon_r = 1 + \frac{\langle \mathbf{M}^2 \rangle - \langle \mathbf{M} \rangle^2}{3\varepsilon_0 V k_B T},
\label{eq:eps_einstein}
\end{equation}
in which $\mathbf{M}$ is the total dipole moment of the simulation box, and $\varepsilon_0$ is the vacuum permittivity.

\subsection{Computational and Experimental Details}
All MD simulations for force field fine-tuning were performed using OpenMM 7.7.0\cite{eastman_openmm_2017}. Simulation boxes were constructed with Packmol using an initial edge length of 40 {\AA}, containing 4000-6000 atoms per system. The exact number of molecules was determined based on the target salt molarity and the pure solvent densities of the components, which resulted in a slight deviation between the target and final molarities. To achieve well-equilibrated systems, a stepwise gradient annealing protocol was employed. Each system was first subjected to a 10-ns NPT simulation at 150 K above the target temperature and 10 bar, using a Monte Carlo barostat. Four sequential 4 ns NPT equilibration stages at 1 bar were then conducted, following a descending temperature gradient of $(T + 100) K, (T + 50) K, (T + 30) K$, and the target temperature $T$. A Nose-Hoover thermostat with a coupling constant of 1 $ps^{-1}$ and a 2-fs time step was used throughout. Non-bonded interactions were truncated at 10 {\AA}, and long-range electrostatic interactions were treated using the Particle Mesh Ewald (PME) method\cite{darden_particle_1993}. Following this equilibration protocol, each force field optimization iteration included a 0.2-ns NVT equilibration run and a 0.4-ns NPT production run.

For property prediction, larger-scale simulations were performed using GROMACS\cite{lindahl_gromacs_2019,berendsen_gromacs_1995}. Equilibrated simulation boxes containing approximately 100,000 atoms were scaled to the target density. Production runs of 35-40 ns in the NVT ensemble with a 2-fs time step were conducted to calculate the self-diffusion coefficients, ionic conductivity, and shear viscosity.
Further details regarding the simulations and experiments are provided in Supporting Information.

\section{Result and Discussion}
\subsection{Model Preparing}

\subsubsection{Performance of Fine-tuned Force Field}
\begin{figure}[H]
\centering
\begin{overpic}[width=0.45\textwidth]{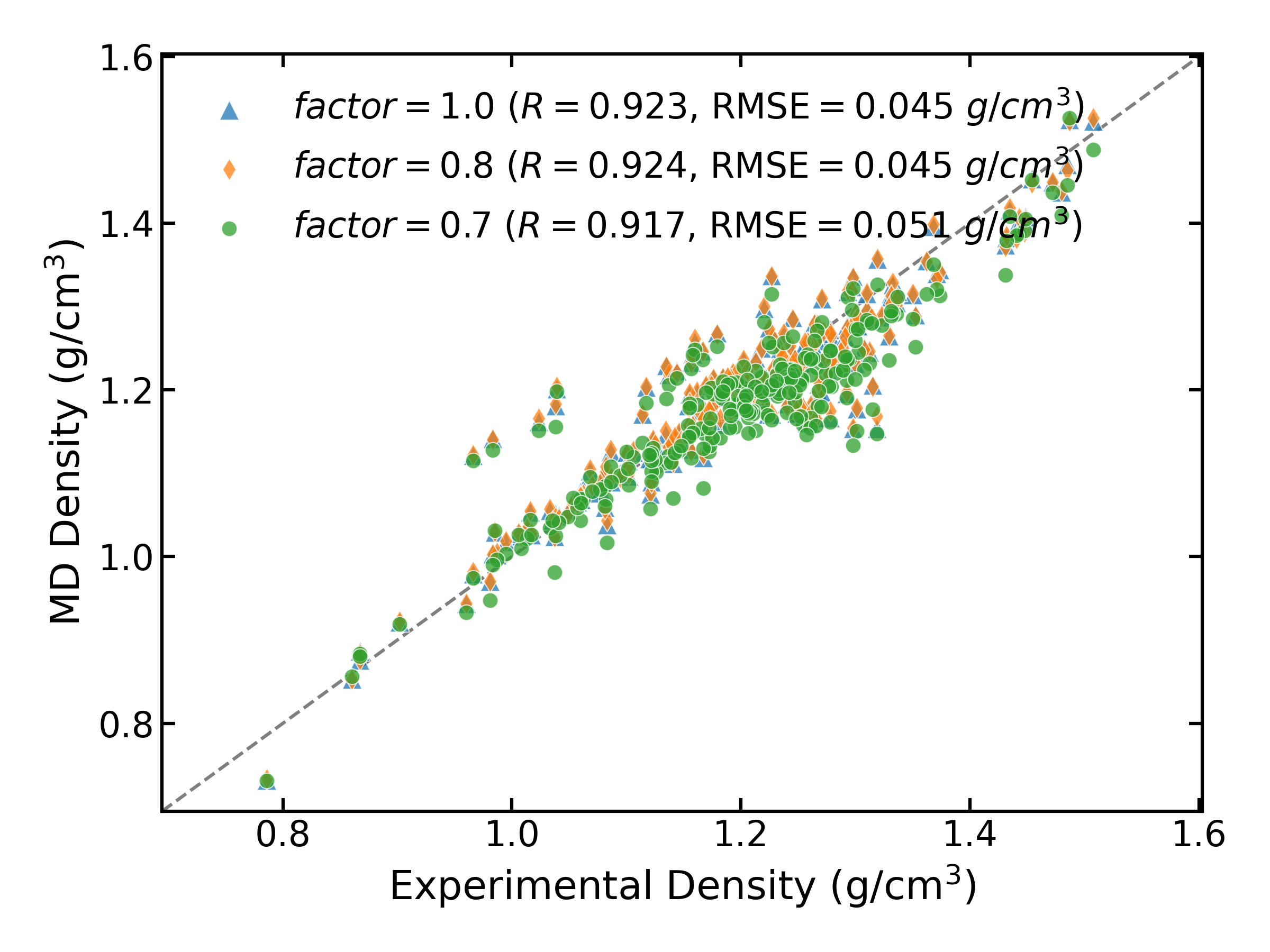}
  \put(0,72){\textbf{(a)}}   
\end{overpic}
\hspace{0.04\textwidth}  
\begin{overpic}[width=0.45\textwidth]{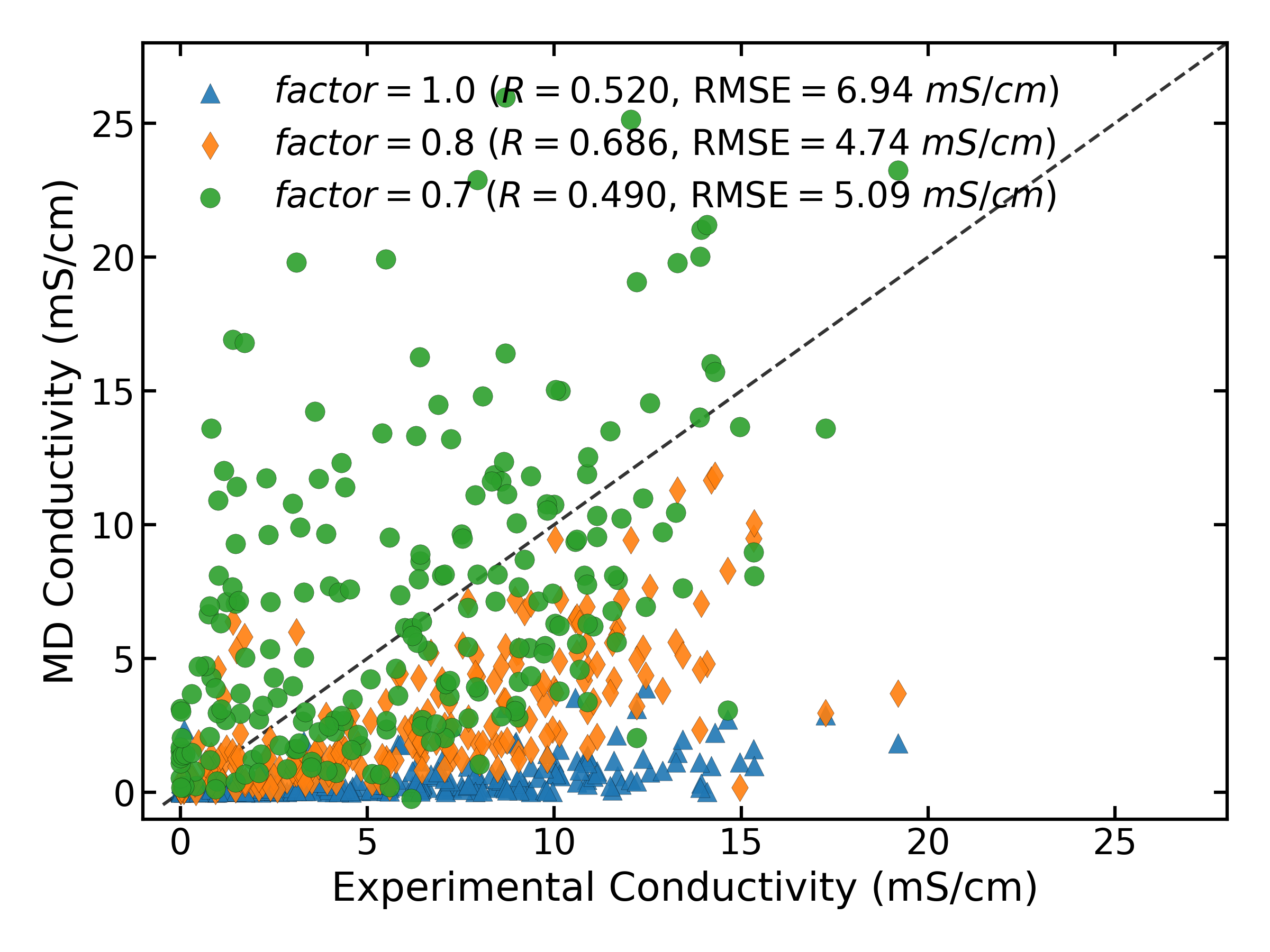}
  \put(1,72){\textbf{(b)}}   
\end{overpic}

\vspace{1em} 
\begin{overpic}[width=0.45\textwidth]{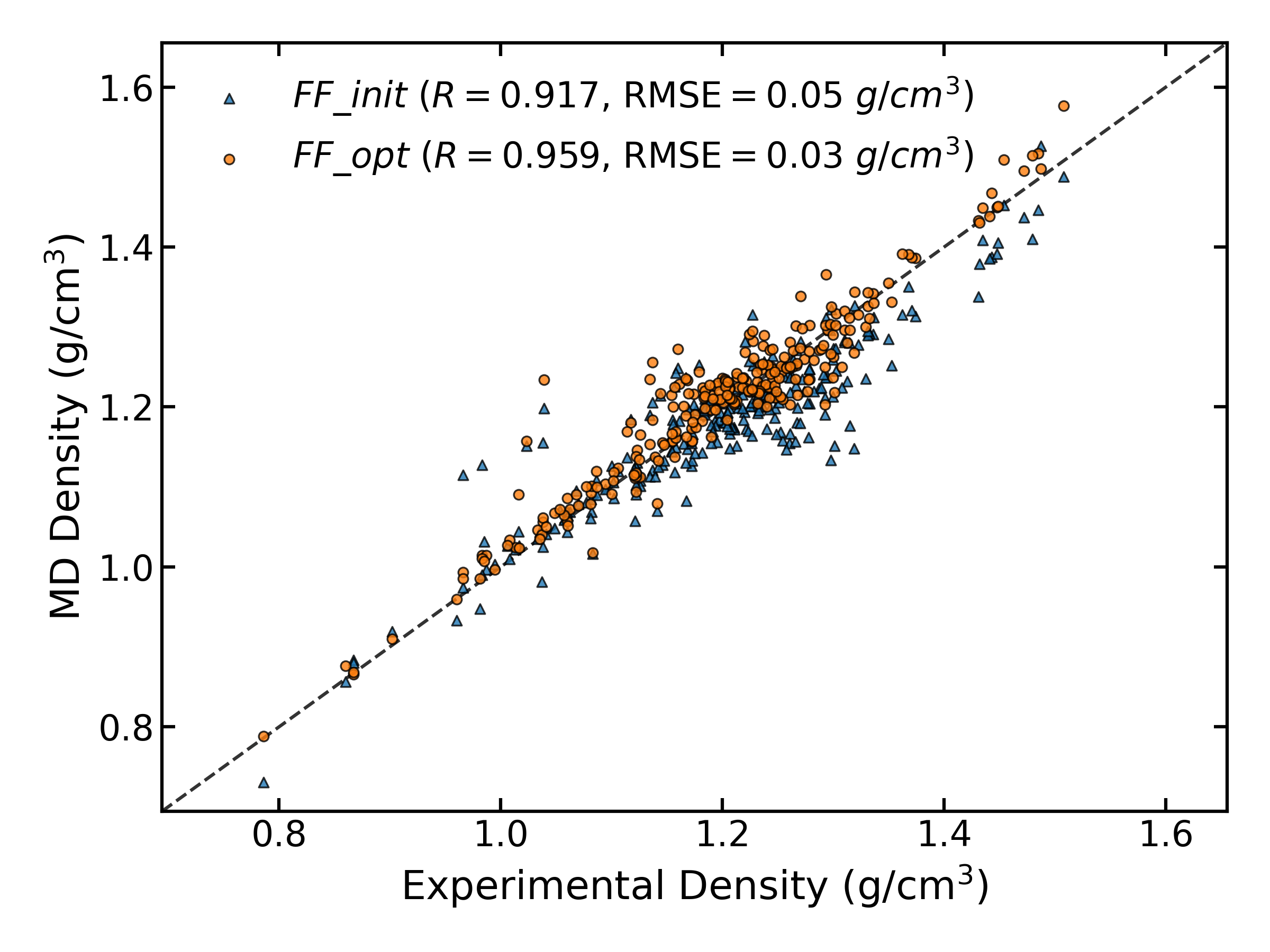}
  \put(1,72){\textbf{(c)}}   
\end{overpic}
\hspace{0.04\textwidth}  
\begin{overpic}[width=0.45\textwidth]{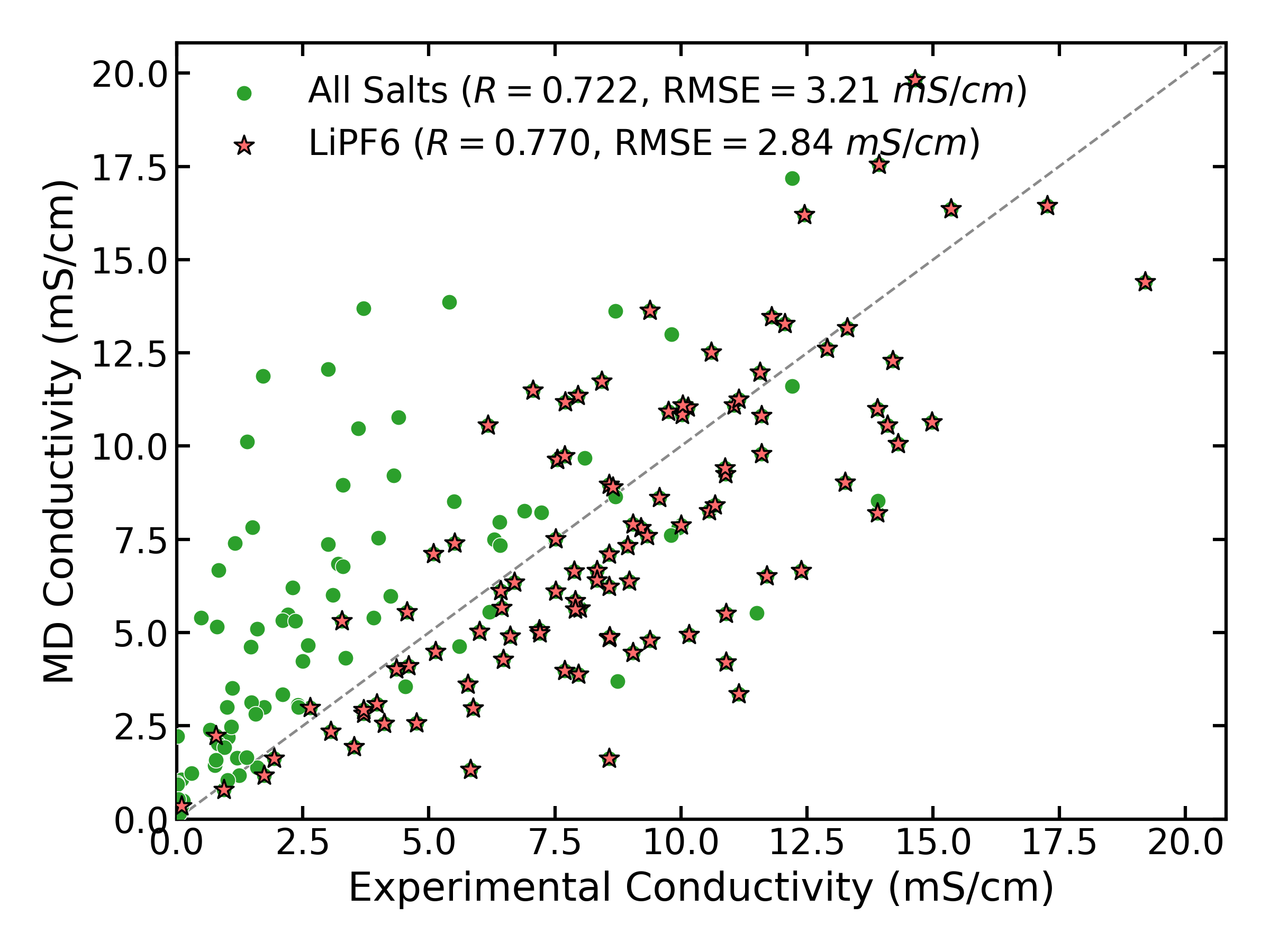}
  \put(1,72){\textbf{(d)}}   
\end{overpic}

\vspace{0.5em} 
\raggedright  
\caption{Density and conductivity prediction performance of models before and after fine-tuning. (a)(b) respectively show the density and Onsager conductivity predicted by the initial OPLS force field with different charge scaling factors \textit{vs} experimental values; (c) shows density predicted by initial and optimized force field within charge scaling factor of 0.7; (d) shows the conductivity predicted by optimized force field within charge scaling factor of 0.7, performance of formulas with \ce{LiPF6} is highlighted by red stars.}
\label{fig:density_conductivity_benchmark}
\end{figure}

Charge scaling is a well-established and computationally efficient approach to implicitly account for electronic polarization effects in classical fixed-charge force fields\cite{cui_influence_2019,liang_charge_2024,fan_charge_2025}. However, arbitrary scaling of charges in pre-parameterized force fields can introduce inconsistencies such as overcorrection of intermolecular interactions. Therefore, before force field fine-tuning, the predictive performance of the initial OPLS-AA force field for density and Onsager ionic conductivity was systematically evaluated using three different charge scaling factors: 1.0 (no scaling), 0.8, and 0.7.
As shown in Figure~\ref{fig:density_conductivity_benchmark}a, charge scaling has a negligible impact on density prediction. All three scaling factors yield excellent agreement with experimental densities, with Pearson correlation coefficients (R) ranging from 0.893 to 0.900 and root-mean-square errors (RMSE) between 0.046 and 0.052 $\mathrm{g/cm^3}$. In stark contrast, conductivity prediction is highly sensitive to the charge scaling factor (Figure~\ref{fig:density_conductivity_benchmark}b). As the scaling factor decreases from 1.0 to 0.7, the predictive accuracy improves dramatically. This improvement arises because unscaled charges overestimate Coulombic interactions between ions, leading to excessive ion aggregation and severe underestimation of ionic conductivity. Compared with 0.8, the scaling factor of 0.7 shows a non-biased prediction of conductivity. Thus, it was selected as the optimal value for further force field fine-tuning and property prediction.
Figures~\ref{fig:density_conductivity_benchmark}c and d compare the predictive performance of the initial force field (INI-FF) and the optimized force field (OPT-FF) using the 0.7 charge scaling factor. For density prediction (Figure~\ref{fig:density_conductivity_benchmark}c), the fine-tuning process significantly improves the accuracy: the Pearson correlation coefficient increases from 0.917 to 0.959, and the RMSE decreases from 0.051 to 0.035 $\mathrm{g/cm^3}$. For ionic conductivity prediction (Figure~\ref{fig:density_conductivity_benchmark}d), the optimized force field achieves an overall RMSE of 3.21 mS/cm across all salt systems. Notably, the commercially dominant \ce{LiPF6}-containing formulations exhibit superior performance with an RMSE of 2.84 mS/cm, consistent with their high representation in our training dataset.

\subsubsection{System-Size and Time Dependence of Transport Properties}
\label{subsec:System_Size_dependence}

\begin{figure}[H]
\centering
\begin{overpic}[width=0.40\textwidth]{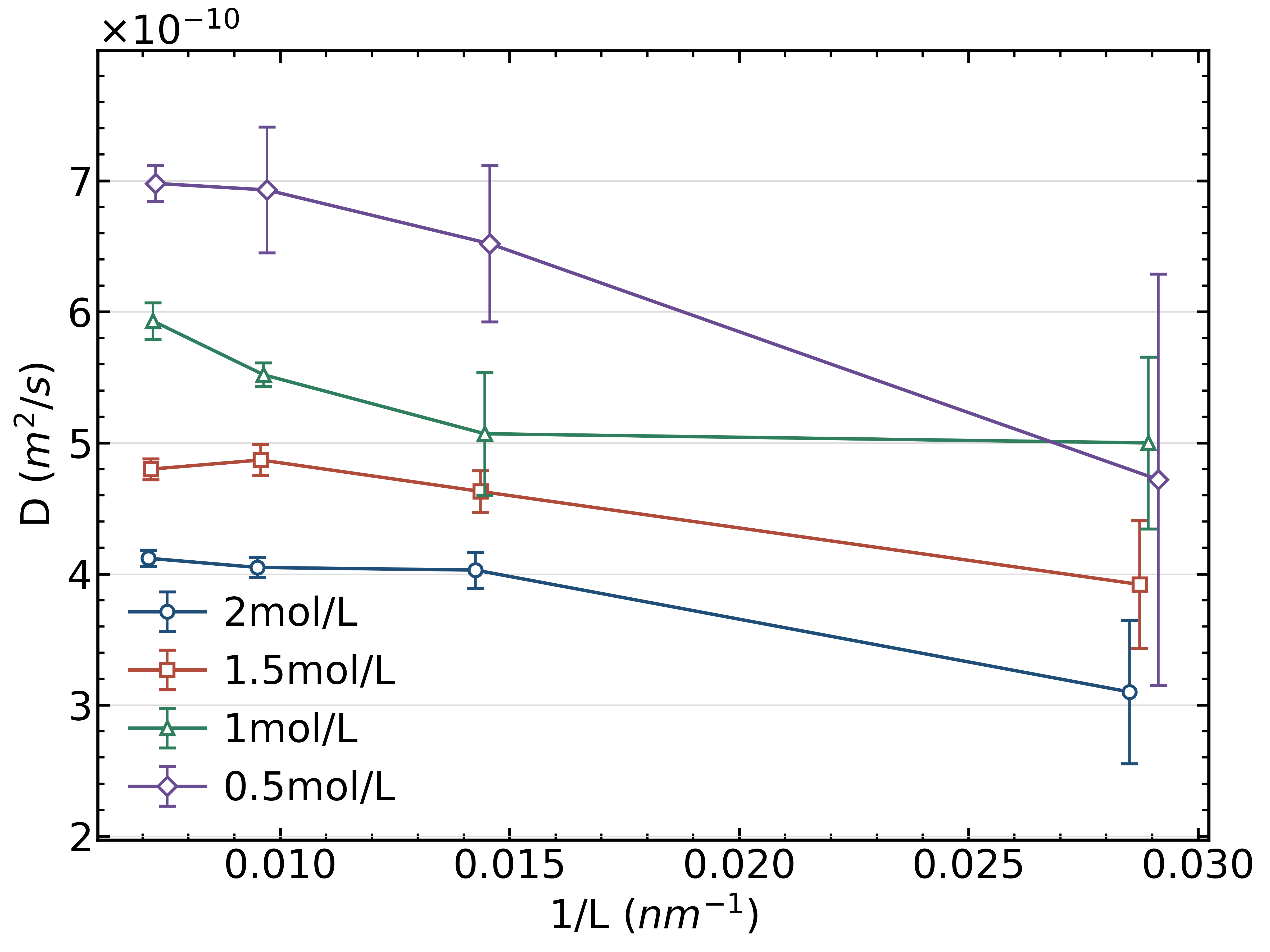}
  \put(-5.0,74){\textbf{(a)}}   
\end{overpic}
\hspace{0.04\textwidth}  
\begin{overpic}[width=0.40\textwidth]{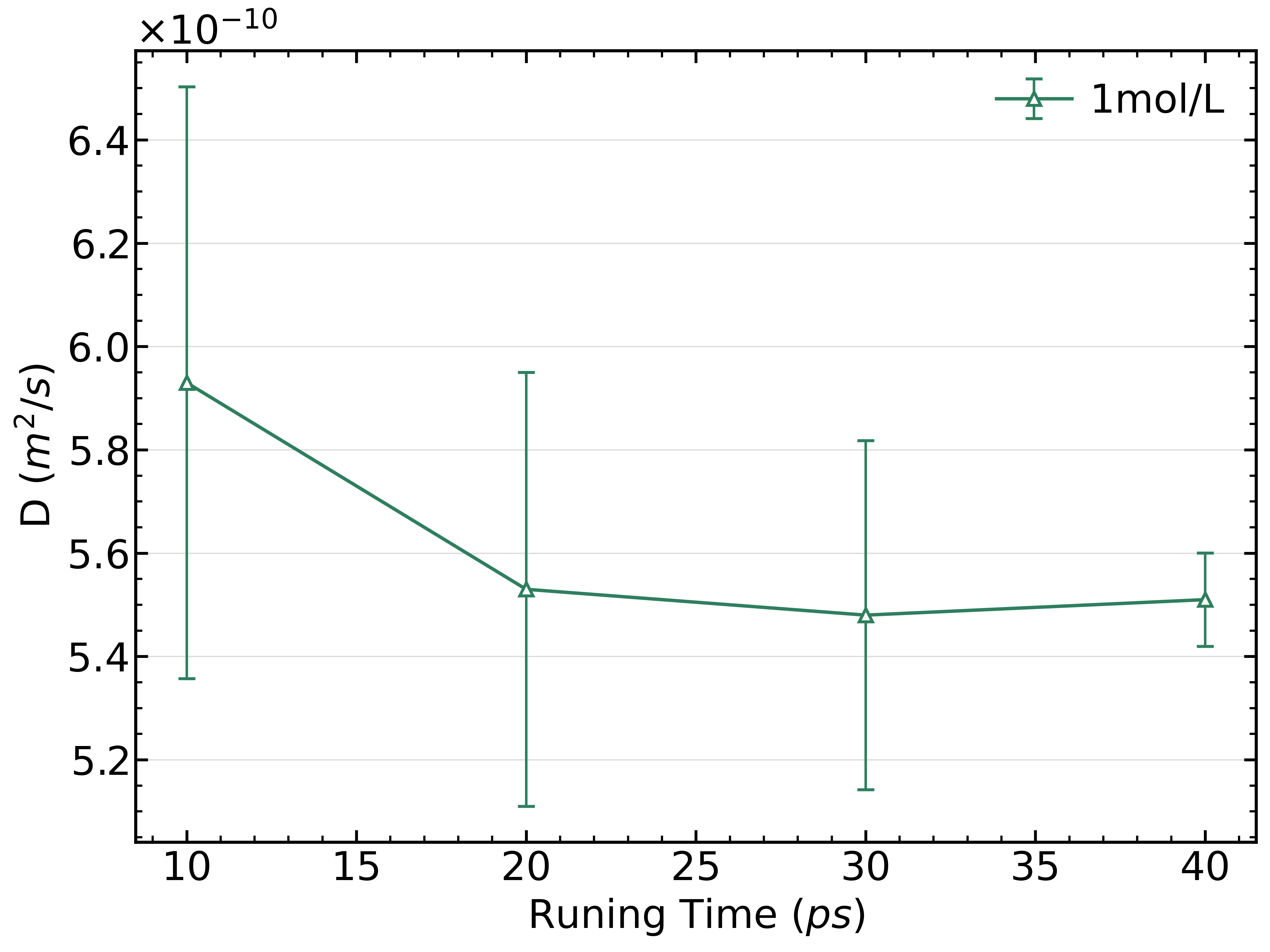}
  \put(-5.0,74){\textbf{(b)}}   
\end{overpic}

\vspace{0.5em} 
\begin{overpic}[width=0.40\textwidth]{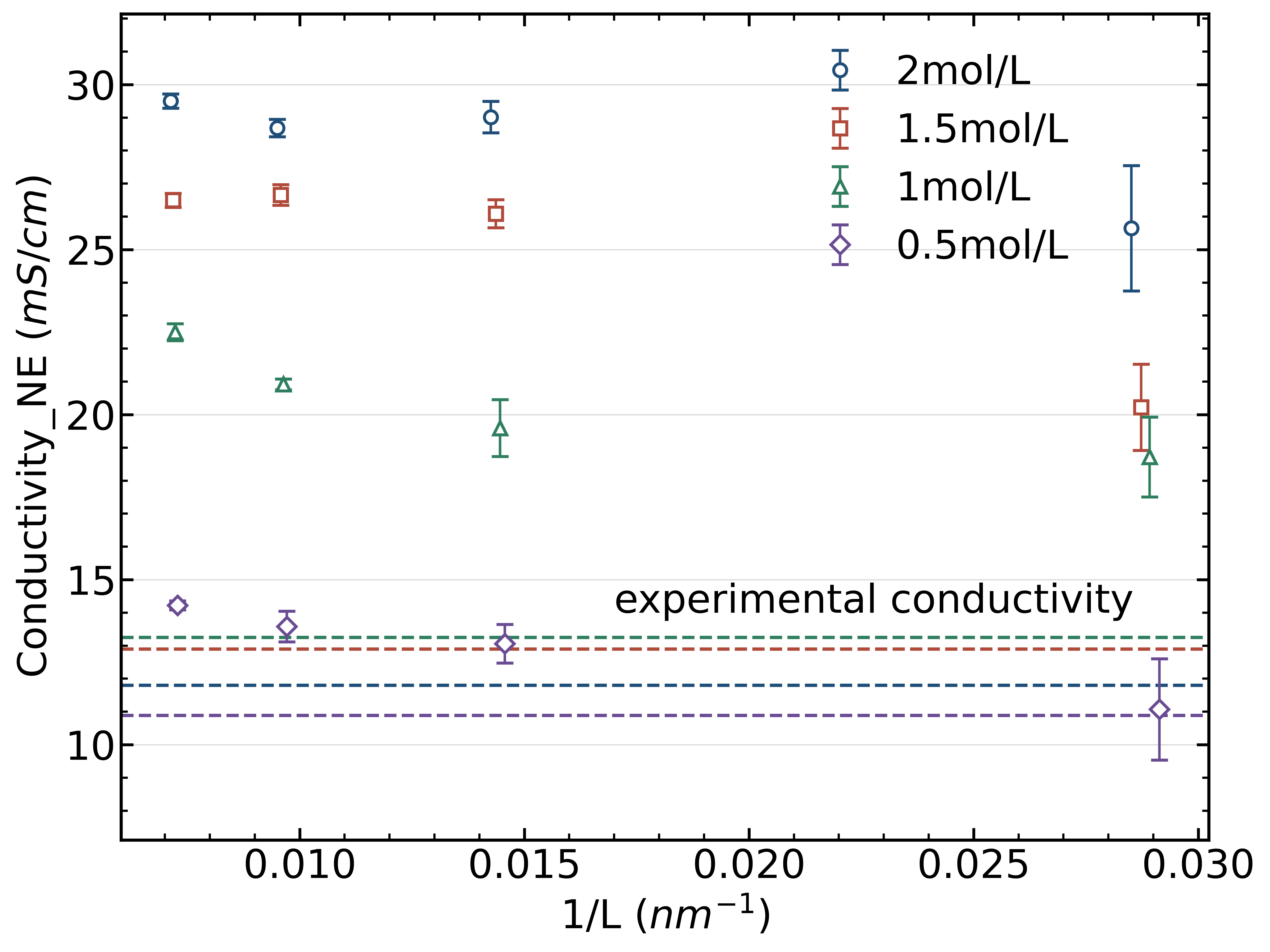}
  \put(-5.0,74){\textbf{(c)}}   
\end{overpic}
\hspace{0.04\textwidth}  
\begin{overpic}[width=0.40\textwidth]{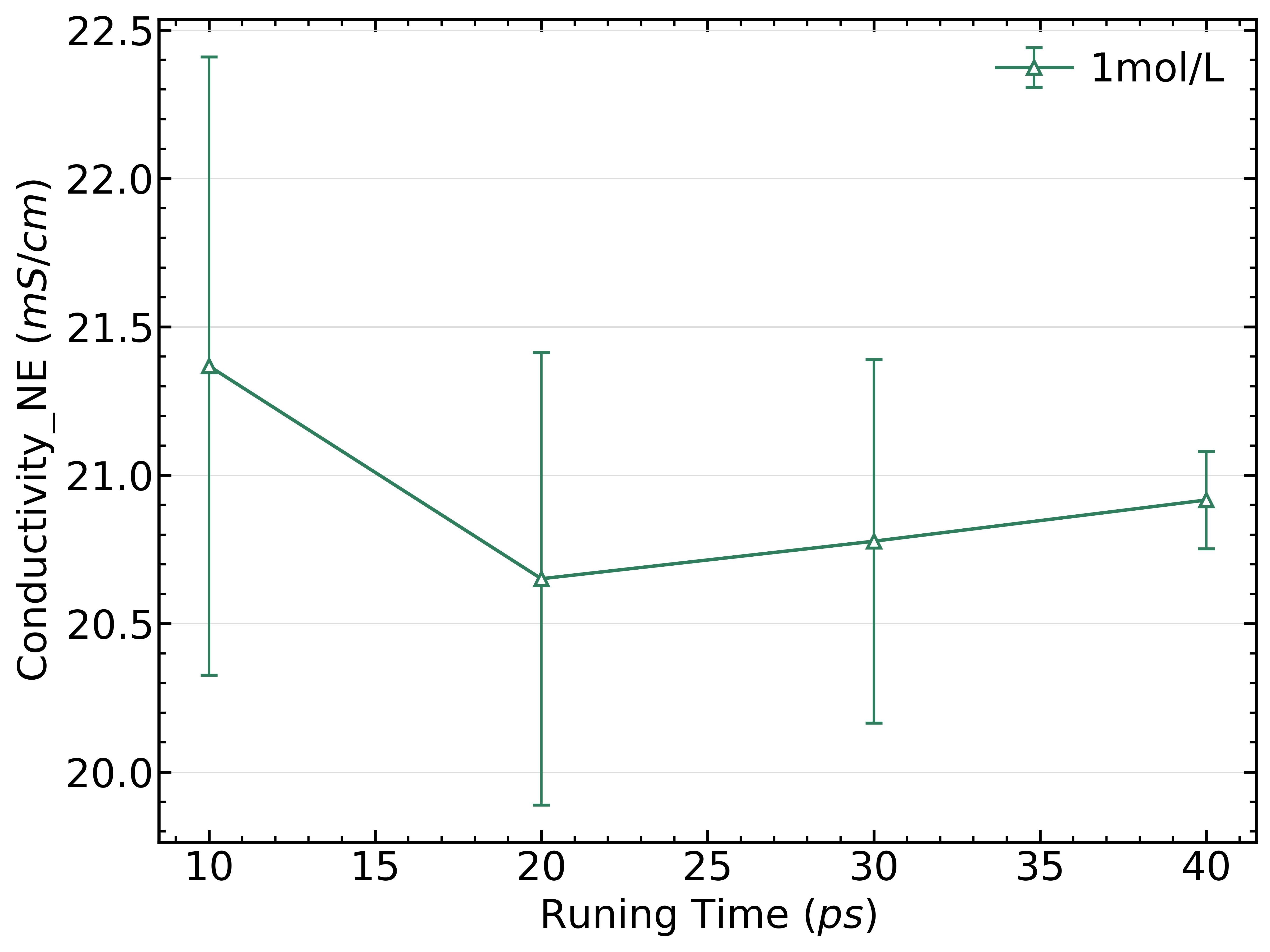}
  \put(-5.0,74){\textbf{(d)}}   
\end{overpic}

\raggedright  
\caption{System-size and time dependence of transport properties at different salt concentrations. (a), (c) show the convergence of $\ce{Li+}$ self-diffusion coefficient and Nernst-Einstein (NE) conductivity as functions of inverse box length, respectively. Dashed lines in (c) denote experimental conductivity values. Blue, orange, green, and purple symbols correspond to salt concentrations of 2.0, 1.5, 1.0, and 0.5 mol/L, respectively. (b), (d) show the time convergence of the corresponding properties at salt concentrations of 1.0 mol/L.}
\label{fig:size_time_convergence}
\end{figure}

To rigorously validate the reliability of our simulation results and quantify finite-size effects introduced by periodic boundary conditions, the system-size and time dependence of key transport properties was comprehensively investigated. A subset of formulations with a fixed solvent composition (EC/DMC = 3:7 by volume) and varying \(\ce{LiPF6}\) concentrations (0.5, 1.0, 1.5, and 2.0 mol/L) was selected from the EDB-1 database for this verification. Base simulation boxes containing $\sim$4,000 atoms were isotropically scaled by factors of 2, 3, and 4 to generate systems with approximately 32,000, 108,000, and 256,000 atoms, respectively. For each system, a 40-ns NVT ensemble production run was performed after equilibration. To assess time convergence, trajectories were segmented into cumulative intervals of 10, 20, 30, and 40 ns. The self-diffusion coefficient of \(\ce{Li+}\), Nernst-Einstein (NE) conductivity, Onsager conductivity, and Green-Kubo viscosity were calculated for each interval and system size. The results are summarized in Figure~\ref{fig:size_time_convergence} and Figure S1 (Supporting Information). As shown in Figures~\ref{fig:size_time_convergence}b and d, the self-diffusion coefficient $D$ and conductivity computed by the NE method reach stable values within 30 ns of simulation time at salt concentrations of 1.0 mol/L, with no statistically significant changes observed beyond this point, and gradually decreasing statistical error. Therefore, the last 10 ns of each production trajectory were used for all subsequent system-size dependence analyses. Figure~\ref{fig:size_time_convergence}a shows that the self-diffusion coefficient of \(\ce{Li+}\) exhibits a clear linear dependence on the inverse box length (\(1/L\)). This behavior is consistent with the hydrodynamic theory of finite-size effects in periodic systems, first systematically characterized by Yeh and Hummer\cite{yeh2004system}. They demonstrated that diffusion coefficients calculated in finite periodic systems are systematically underestimated due to long-range hydrodynamic self-interactions between a particle and its periodic images. As expected, larger systems exhibit smaller statistical uncertainties due to improved configurational sampling. The NE conductivity, derived directly from the self-diffusion coefficients of cations and anions, displays an identical \(1/L\) scaling behavior (Figure~\ref{fig:size_time_convergence}c), as it inherits the system-size dependence of the underlying diffusion coefficients. However, NE conductivity significantly overestimates experimental values at higher salt concentrations. This discrepancy arises from the fundamental limitation of the NE approximation, which neglects correlated ion motion. In concentrated electrolytes, strong Coulombic interactions lead to correlated diffusion of cations and anions, which reduces net charge transport but is not accounted for in the NE formalism. In contrast, the Onsager conductivity, which explicitly incorporates cross-correlations between all ionic species, shows excellent quantitative agreement with experimental values across all salt concentrations (Figure S1a). However, the Onsager method exhibits larger statistical fluctuations compared to the NE method, as it requires accurate calculation of cross-correlation functions between different ion pairs. Notably, the statistical error of Onsager conductivity does not decrease monotonically with increasing system size, which can be attributed to the increased number of correlated ion pairs in larger systems that contribute to noise in the correlation functions. Shear viscosity was calculated using the time-decomposition Green-Kubo formalism proposed by Zhang et al., which improves reliability by mitigating the effects of long-time tails in the stress autocorrelation function. As shown in Figure S1c (Supporting Information), viscosity shows no significant system-size dependence within the range of box sizes investigated. This observation is in excellent agreement with the findings of Yeh and Hummer\cite{yeh2004system}, who demonstrated that shear viscosity is essentially independent of system size for homogeneous fluids under periodic boundary conditions, in stark contrast to diffusion coefficients. Based on these comprehensive convergence studies, a standard simulation protocol was selected for all subsequent calculations: a box size containing approximately 100,000 atoms and a 35-ns NVT production run. This choice balances computational efficiency with statistical accuracy and minimizes finite-size effects on the calculated transport properties.

\subsubsection{Performance of the Tianqiong platform}
\begin{figure}[H]
\centering
\begin{overpic}[width=0.80\textwidth]{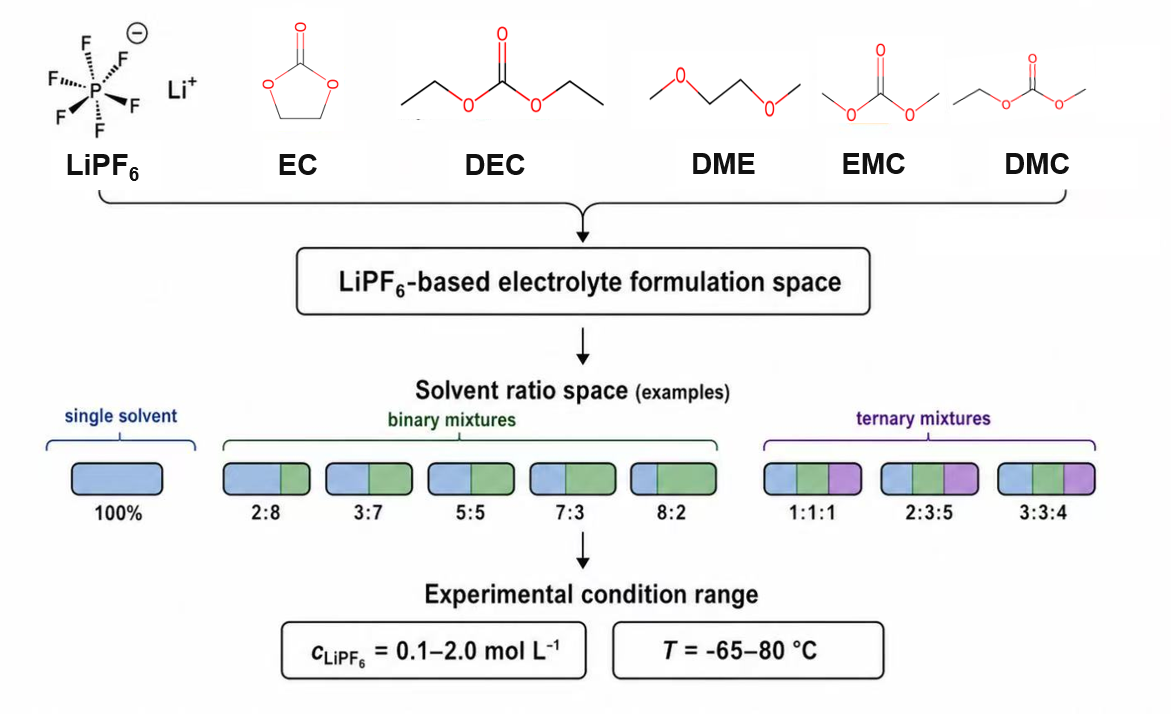}
\put(-4,62){\textbf{(a)}}
\end{overpic}
\vspace{0.04\textwidth}
\begin{overpic}[width=0.90\textwidth]{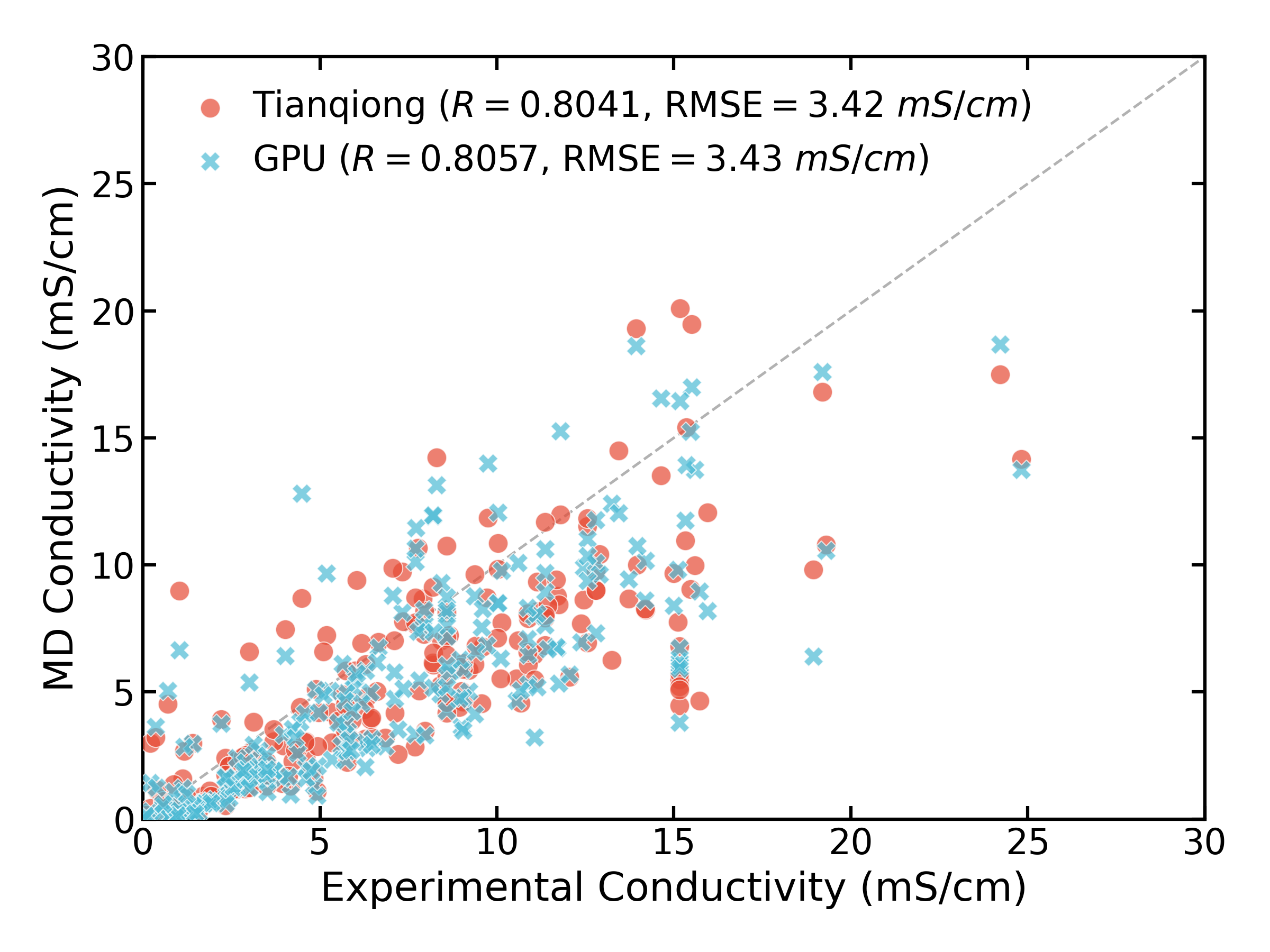}
\put(1,72){\textbf{(b)}}
\end{overpic}
\caption{Chemical space of screened electrolyte formulations from the EDB-1 database and cross-platform comparison of ionic conductivity. (a) Chemical space of the test 231 electrolyte formulations, covering \ce{LiPF6}-based systems with common solvents (\ce{DEC}, \ce{EC}, \ce{DMC}, \ce{DME}, \ce{EMC}); (b) Parity plot of MD-predicted versus experimental ionic conductivity for selected \ce{LiPF6}-based electrolytes, with Pearson correlation coefficients ($R$) and root-mean-square errors (RMSE) presented for the Tianqiong platform and a selected GPU (RTX 4090).}

\label{fig:accuracy_comparsion_with_gpu}
\end{figure}

Large system and long-time molecular dynamics (MD) simulations represent the primary computational bottleneck in high-throughput electrolyte screening workflows. To address this challenge, we employed the Tianqiong platform for electrolyte MD simulations. The predictive accuracy was compared with results in seleted GPU nodes. Figure~\ref{fig:accuracy_comparsion_with_gpu}a briefly presents the chemical space of the tested electrolyte formulations, i.e., a set of \ce{LiPF6}-based electrolyte systems, and Figure~\ref{fig:accuracy_comparsion_with_gpu}b compares the accuracy of ionic conductivity results on the two platforms using our fine-tuned force field.  The Tianqiong platform yields a Pearson correlation coefficient (R) of 0.8041 and a root-mean-square error (RMSE) of 3.42 mS/cm, while the GPU platform produces nearly identical results with \(R=0.8057\) and RMSE=3.43 mS/cm. The negligible difference between the two datasets confirms that Tianqiong fully preserves the physical fidelity of the MD simulations and introduces no systematic bias in the calculated transport properties. For other transport properties, including self-diffusion coefficients and shear viscosity, for which comprehensive experimental data are not available, quantitative agreement within statistical uncertainty between the two platforms was observed. These additional comparisons are provided in Supporting Information (Figure S2). The following high-throughput electrolyte simulations were performed on the Tianqiong platform due to its high accuracy and efficiency.

\subsection{Electrolyte Property dataset}
Leveraging the exceptional computational efficiency of the Tianqiong platform, over 10,000 large-scale MD simulations were performed to construct a comprehensive electrolyte property dataset covering both single-salt and dual-salt formulations. The single-salt systems were extracted from the EDB-1 database with available experimental ionic conductivity data, serving as the benchmark set for force field accuracy validation. For the dual-salt systems, the formulation space was extended by introducing a second functional lithium salt (including \ce{LiDFOB}, \ce{LiBOB}, \ce{LiTFSI}, \ce{LiFSI}, \ce{LiBF4}, and \ce{LiClO4}) into the baseline \ce{LiPF6} electrolyte, aiming to expand the property landscape and explore novel multi-component formulation candidates. For each formulation, five key physicochemical properties were calculated: density, ionic conductivity, lithium-ion self-diffusion coefficient, shear viscosity, and relative dielectric constant. This high-throughput computational screening effort generated over 50,000 data points, providing a rich data resource for electrolyte molecular design and force field generality validation.

\subsubsection{Predictive Performance of Ionic Conductivity}
\begin{figure}[H]
\centering
\begin{overpic}[width=0.45\textwidth]{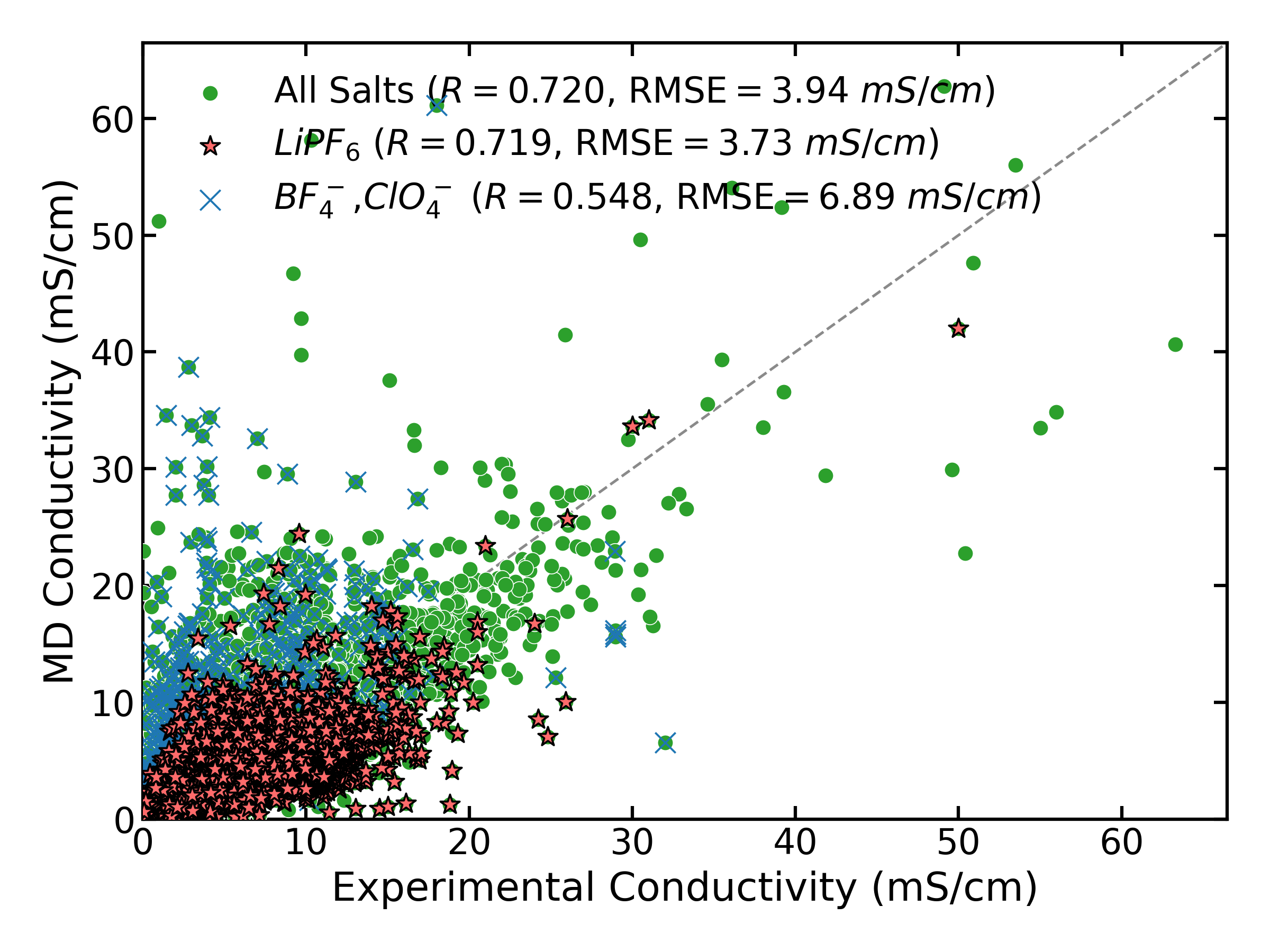}
\put(0,72){\textbf{(a)}}
\end{overpic}
\hspace{0.04\textwidth}
\begin{overpic}[width=0.45\textwidth]{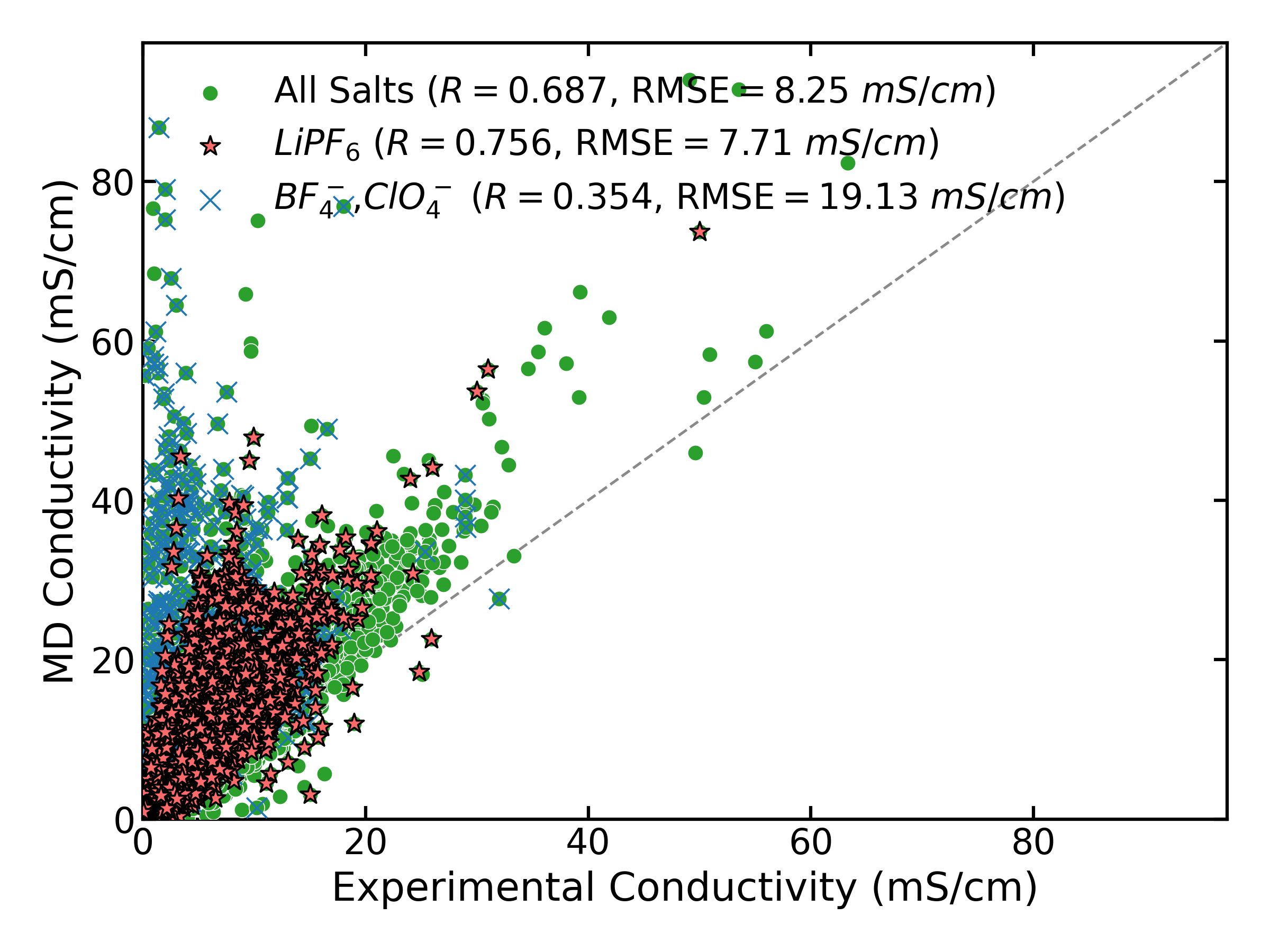}
\put(1,72){\textbf{(b)}}
\end{overpic}
\caption{Parity plots of MD-predicted vs experimental ionic conductivity for single-salt electrolyte formulations screened from the EDB-1 database. (a) Conductivity calculated using the Onsager transport theory; (b) Conductivity calculated using the Nernst-Einstein (NE) approximation. Green circles represent all salt systems, red stars denote \ce{LiPF6}-based formulations, and blue crosses indicate systems containing \(\ce{BF4-}\) or \(\ce{ClO4-}\) anions. Pearson correlation coefficients (R) and root-mean-square errors (RMSE) are provided for each subset.}
\label{fig:high_throughput_comparison}
\end{figure}

The predictive performance of our optimized force field was first evaluated on the single-salt benchmark set from the EDB-1 database, as illustrated in Figure~\ref{fig:high_throughput_comparison}. Two widely used conductivity calculation methods were compared: the Onsager transport theory, which explicitly accounts for ionic cross-correlations, and the Nernst-Einstein (NE) approximation, which neglects correlated ion motion.

As shown in Figure~\ref{fig:high_throughput_comparison}a, the Onsager method achieves an overall Pearson correlation coefficient (R) of 0.720 and a root-mean-square error (RMSE) of 3.94 mS/cm across all electrolyte systems. In contrast, the NE approximation (Figure~\ref{fig:high_throughput_comparison}b) yields a lower overall $R$ of 0.687 and a significantly higher RMSE of 8.25 mS/cm. The strength of the Onsager method lies in its explicit treatment of cation–anion correlated diffusion--a feature that is especially critical in concentrated electrolytes, where strong Coulombic interactions dominate transport behavior. The NE approximation systematically overestimates conductivity at high values, as it fails to account for the reduction in net charge transport caused by ion pairing and collective ionic motion. Notably, for commercially dominant \ce{LiPF6}-based electrolytes, the NE method yields a marginally higher Pearson correlation (
$R$=0.756 vs. 0.719) than the Onsager method. However, the Onsager method provides more accurate absolute values with considerably smaller systematic bias, highlighting its superior physical rigor for quantitative predictions across diverse electrolyte systems.
The performance across different anion chemistries was further analyzed. Systems containing \(\ce{BF4-}\) or \(\ce{ClO4-}\) anions show substantially poorer predictive accuracy, with $R$ values of only 0.548 (Onsager) and 0.354 (NE). Although the Lennard-Jones parameters for these anions were optimized during force-field fine-tuning, their use led to severe ion aggregation and compromised stability in extended, large-scale simulations. To maintain numerical reliability across the entire high-throughput formulation library, the original OPLS-AA parameters were therefore retained for these two species in all production runs. This retention, however, comes at the cost of a less accurate description of ion–solvent interactions, which ultimately degrades the predictive performance.

\subsubsection{High-Dimensional Property Landscape Visualized by t-SNE}
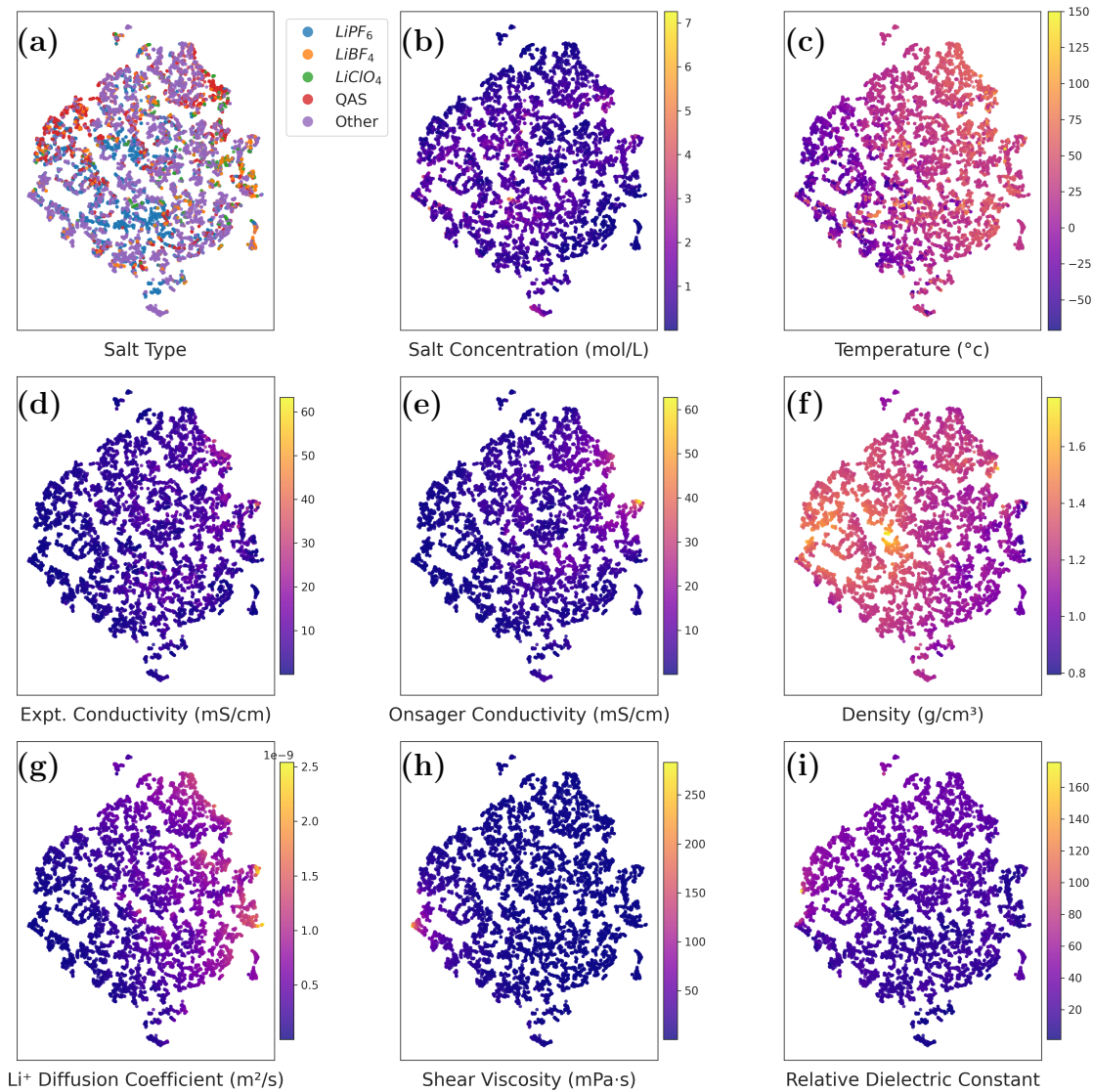
\begin{figure}[H]
\centering
\begin{overpic}[width=0.9\textwidth]{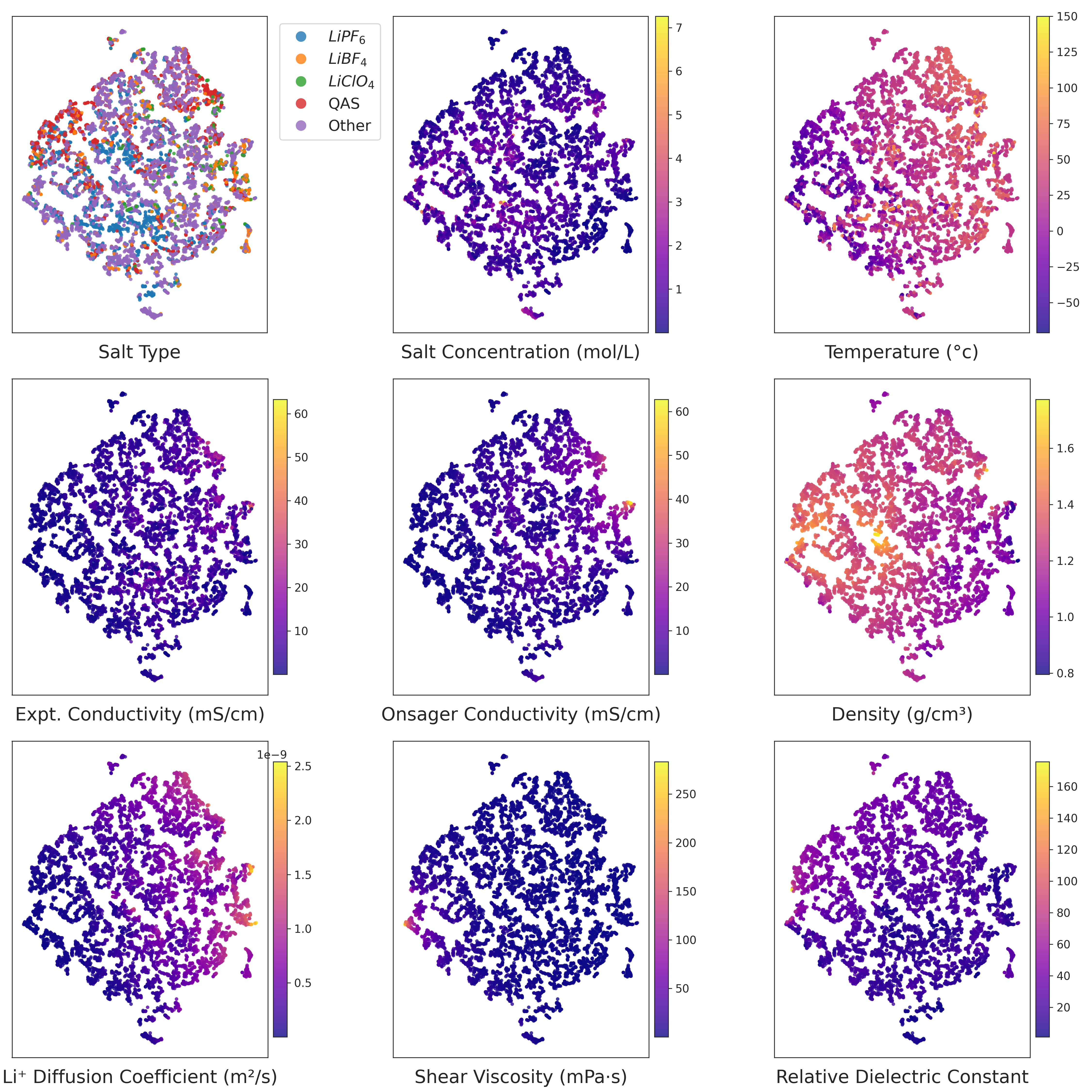}
    \put(1, 95){\textbf{(a)}}   
    \put(36, 95){\textbf{(b)}}  
    \put(71, 95){\textbf{(c)}}  
    
    \put(1, 62){\textbf{(d)}}   
    \put(36, 62){\textbf{(e)}}  
    \put(71, 62){\textbf{(f)}}  
    
    \put(1, 29){\textbf{(g)}}   
    \put(36, 29){\textbf{(h)}}  
    \put(71, 29){\textbf{(i)}}  
\end{overpic}
\caption{t-SNE 2D embedding of the single-salt electrolyte property dataset. All embeddings are generated using five core physicochemical properties: density, Onsager ionic conductivity, \ce{Li+} self-diffusion coefficient, Green-Kubo shear viscosity, and relative dielectric constant. Panels are colored by (a) salt type, with \ce{LiPF6}, \ce{LiBF4}, \ce{LiClO4}, quaternary ammonium salts (QAS), and other lithium salts highlighted; (b) salt concentration; (c) temperature; (d) experimental ionic conductivity; (e) MD-predicted Onsager ionic conductivity; (f) density; (g) \ce{Li+} self-diffusion coefficient; (h) shear viscosity; and (i) relative dielectric constant. Green-Kubo viscosity calculations frequently fail to reach full convergence for systems at low temperatures and high salt concentrations. As the t-SNE embedding requires complete data for all five properties per sample, formulations with non-convergent viscosity trajectories were excluded entirely, yielding a final set of 7868 valid single-salt electrolyte formulations out of 8077 total candidates.}

\label{fig:tsne_single_salt}
\end{figure}

Beyond single-property validation, stochastic neighbor embedding (t-SNE)\cite{maaten_visualizing_2008} was used to visualize the 5-dimensional property space (density, Onsager conductivity, \ce{Li+} diffusivity, shear viscosity, and dielectric constant) of the dataset, uncovering hidden formulation-property correlations. The algorithm nonlinearly maps high-dimensional descriptors to a 2D plane while preserving local similarity: adjacent points correspond to electrolytes with comparable overall physicochemical profiles.
Figure~\ref{fig:tsne_single_salt} presents the t-SNE embedding of the single-salt benchmark set. Colored by salt type (Figure~\ref{fig:tsne_single_salt}a), the 15 salt systems show partial regional aggregation rather than fully isolated clusters, indicating that anion chemistry is a key modulator while solvent composition and concentration also contribute to overlapping property spaces. The unoptimized \(\ce{BF4-}\) and \(\ce{ClO4-}\) systems are scattered across regions, consistent with their larger prediction deviations.

Salt concentration exhibits a clear continuous gradient (Figure~\ref{fig:tsne_single_salt}b): low-concentration formulations (0-2 mol/L) dominate the central-right region, while high-concentration systems ($>4$ mol/L) aggregate in the left and upper-left, confirming concentration as the dominant factor governing overall electrolyte properties. Temperature also shows a gradual spatial distribution (Figure~\ref{fig:tsne_single_salt}c), with low- and high-temperature systems clustering in the lower-left and lower-right, respectively, reflecting thermal modulation of transport properties.

Color-mapped property distributions (Figures~\ref{fig:tsne_single_salt}d-i) reveal clear correlations among the computed features. Experimental and Onsager-predicted conductivity share highly consistent spatial patterns (Figures~\ref{fig:tsne_single_salt}d-e), globally verifying the reliability of the optimized force field. Density and viscosity follow similar gradients with high values in the left high-concentration region (Figures~\ref{fig:tsne_single_salt}f, h), while \ce{Li+} diffusivity shows the opposite trend (Figure~\ref{fig:tsne_single_salt}g). This positive density-viscosity correlation and the inverse relation between diffusion and viscosity agree well with established physical expectations. In particular, the latter is in good accordance with the Stokes-Einstein relation, confirming the physical self-consistency of the simulations. The relative dielectric constant distributes relatively uniformly (Figure~\ref{fig:tsne_single_salt}i), suggesting it is mainly determined by solvent components and less sensitive to salt type.

Further, dual-salt systems comprising \ce{LiPF6} as the base salt and six functional lithium salts as additives were constructed. The corresponding t-SNE embeddings are provided in Figure S3 (Supporting Information). Dual-salt formulations partially overlap with the single-salt property space while extending into uncovered regions, verifying the dual-salt strategy as an effective approach to tune comprehensive electrolyte properties\cite{zhao_strong_2023,zheng_lithium_2020,du_engineering_2024}. Sulfonimide-based (\ce{LiTFSI}/\ce{LiFSI}) dual-salt systems maintain a favorable balance of moderate-to-high conductivity and moderate viscosity, while borate-based (\ce{LiBOB}/\ce{LiDFOB}) systems tend to distribute in higher density and viscosity regions, attributable to their larger anionic volume and stronger intermolecular interactions with solvents\cite{zhao_strong_2023,zheng_lithium_2020,du_engineering_2024}.

In addition, the probability density distributions of all predicted properties for both single-salt and dual-salt systems are compiled in Figure S4 (Supporting Information). These univariate distributions offer a direct statistical perspective of the full formulation library, complementing the high-dimensional t-SNE analysis and demonstrating the broad property coverage of the constructed electrolyte dataset.

\section{Conclusion}
In this work, an accurate and transferable OPLS-AA force field for ionic electrolytes has been developed through an automated differentiable parameterization workflow, which shows excellent performance in simulating over 10,000 electrolyte formulations that encompass 67 solvents and 15 lithium salts.
Our primary strategies include: 
(1) topology-guided atom typification that greatly reduces parameter redundancy and condenses Lennard-Jones parameters to 136;
(2) a dual-property validation strategy---density fitting for structural and thermodynamic properties, and conductivity for transport properties---that couples DMFF with experimental references to ensure balanced accuracy across structural and transport properties;
(3) a rigorous convergence study confirming the $1/L$ scaling of diffusion coefficients, and thereby establishing a standardized MD protocol for high-throughput simulations.

Based on the delicately optimized force field and the computational protocol, a large MD dataset of LIBs electrolytes has been constructed using the Tianqiong platform, which contains 50,000 formulation-level data points for the five key physical properties including density, dielectric constant, viscosity, diffusion coefficient, and ionic conductivity. t-SNE-based high-dimensional property landscape visualization reveals the continuous gradient of electrolyte properties with salt concentration and temperature, demonstrates the partial clustering characteristics of different salt systems, and verifies the internal consistency of multiple physicochemical properties from a global perspective. The methodologies and data resources developed here are expected to significantly accelerate the rational design and discovery of advanced electrolyte systems, especially with the help of advanced AI methodologies.

Limitations come from unoptimized atom-type parameters for specific anions (e.g., \ce{B} and \ce{F} in  \(\ce{BF4-}\), \ce{Cl} and \ce{O} in \(\ce{ClO4-}\)), and the inherent constraints of the adopted fixed-charge model. Future work will expand the training set to more emerging functional anions during parameter optimization, develop polarizable force field models for better description of strong ionic interactions, and extend the dataset to broader conditions. Combined with machine learning approaches, these methodological and data resources will enable targeted, high-throughput inverse design of application-oriented high-performance electrolytes, further propelling the paradigm shift of electrolyte research from empirical trial-and-error to data-driven rational discovery.

\section{Acknowledgments}
This work was supported by the Advanced Materials-National Science and Technology Major Project (grant no. 2025ZD0619800). The authors gratefully acknowledge the engineering team at Shanghai Smart Logic Technology for their contributions to data generation and curation, the Shanghai Academy of AI for Science for their collaborative support, and the Changjiang 3D Scientific Computing Center for providing computational resources.

\setcounter{figure}{0}
\renewcommand{\thefigure}{S\arabic{figure}}
\setcounter{table}{0}
\renewcommand{\thetable}{S\arabic{table}}
\renewcommand{\theequation}{S\arabic{equation}}
\section*{Supporting Information}
\setcounter{section}{0}
\renewcommand{\thesection}{S\arabic{section}}

\section{Computational Details}
\subsection{Electrolyte Screening}
The specific screening criteria are described below: (1) Neutral molecules appearing fewer than 10 times and salts appearing fewer than 20 times were excluded. (2) All sodium salts and imidazolium salts were discarded as they are irrelevant to lithium-ion battery systems. Quaternary ammonium salts, widely used as electrolyte additives or co-salts, were retained. (3) Besides bis(oxalato)borate (\ce{BOB-}), difluoro(oxalato)borate (\ce{DFOB-}), and tetrafluoroborate (\ce{BF4-}), all other boron-containing salts were excluded due to the scarcity of validated initial force field parameters and poor commercial accessibility. (4) Besides hexafluorophosphate (\ce{PF6-}), all other organic phosphorus-containing salts were removed for parameter unavailability and limited short-term commercial supply. (5) Imidazolium-based anions were eliminated from the database. (6) Molecules containing reactive hydrogen moieties, specifically carboxylic acid ester (\ce{-COO-}) and sulfonamide (\ce{-SO2NH-}) groups, were excluded as they exhibit poor chemical stability under lithium battery operating conditions.

\subsection{Experimental Measurement}
\begin{itemize}

\item \textbf{Electrolyte Samples Preparing}: Battery-grade solvents (ethylene carbonate (EC, 99.9\%), Dimethyl carbonate (DMC,99.9\%), methyl ethyl carbonate (EMC, 99.9\%), Diethyl carbonate (DEC, 99.9\%), etc) were purchased from Suzhou Dodo Chemical Technology Co., Ltd. (China) and used without further purification. 
And lithium salts (Lithium Bis(fluorosulfonyl)imide (\ce{LiFSI}, $≥$99\%)，Lithium bis(trifluoromethanesulfonyl)imide (\ce{LiTFSI}, $≥$99\%), lithium difluoro(oxalato)borate ( \ce{LiBOB}, $≥$99\%), lithium hexafluorophosphate (\ce{LiPF6}, $≥$99\%), etc) were obtained from Shanghai Aladdin Biochemical Technology Co., Ltd. (China).
All electrolyte samples were made and stored in an argon-filled glovebox with \ce{O2} and \ce{H2O} levels $<$ 0.1 ppm.

\item \textbf{Density measurement}: Densities of all electrolyte samples were determined at 298.15 K using a Youyunpu YP-DS handheld digital densitometer, which provides a measurement precision of $\pm 0.0001\, \mathrm{g/cm^3}$. Each sample was measured three consecutive times under identical conditions, and the arithmetic mean was calculated and reported.A total of 21 solvent molecules and 15 salts were selected to prepare single-salt binary-solvent 2 formulations, covering all 68 optimized atom types. The salt concentrations were set to 0.3 $\mathrm{M}$, 0.5 $\mathrm{M}$, and 1.0 $\mathrm{M}$, and the solvent volume ratios were 5:5 or 3:7 with EC/DMC as the base solvent.

\item \textbf{Conductivity validation}: Independent conductivity data were extracted from the EDB-1 database, which were not used in the force field fitting process. These formulations included major commercial electrolyte components such as \ce{LiPF6}/\ce{EC}/\ce{DMC}, \ce{LiFSI}/\ce{EMC}, and \ce{LiTFSI}/\ce{GBL}, covering a conductivity range of 1–20 mS/cm.
\end{itemize}
\subsection{ Force Field Parameter Optimization}
\begin{itemize}
\item \textbf{Initial force field construction}: Most OPLS-AA force field parameters were extracted from the LigParGen\cite{dodda_114cm1a-lbcc_2017,dodda_ligpargen_2017,jorgensen_potential_2005} web server using the $1.14\times$CM1A charge model. For molecules not supported by LigParGen (e.g., nitriles, borates, and most anions), parameters were collected from existing literature\cite{sambasivarao_development_2009,doherty_revisiting_2017,palluzzi_insight_2024,rana_fast_2025,mohapatra_effect_2022,klinov_transferable_2021}. For extremely rare molecules, specifically compound with SMILES C1COB(OCCOB2OCCCO2)OC1, atomic charges were obtained from MP2/cc-pVTZ ChelpG charge by Gaussian, and LJ parameters were transferred from chemically similar atoms.
\item \textbf{Atom typification}: All atoms in the selected electrolyte components were classified based on their chemical environments, defined by the number, element type, and hybridization state of their first-nearest neighbors. Atoms with identical chemical environments were assigned the same atom type. This process reduced the original OPLS atom types to 78 unique types, in which a small subset of molecules that are not commercially available but contain unique atom types. Two distinct parameterization strategies were adopted for these atom types, through which atom types were further condensed to 68 types. One strategy is for Silicon-containing species, we directly adopted the original OPLS-AA parameters for all silicon-containing atoms without further optimization. Another is for Fluorinated carbon species, we implemented a manual atom type mapping approach based on chemical similarity. 
\item \textbf{Grouped iterative optimization}: The LJ parameters were optimized using a grouped cyclic optimization strategy to ensure broad applicability across all electrolyte systems. Initially, the entire set of electrolyte formulations was divided into batches of eight formulations each. For each batch, NPT MD simulations were performed using OpenMM\cite{eastman_openmm_2017} accelerated by eight GPU accelerators with the current force field parameters. Subsequently, the loss function was calculated based on the deviation between simulated and experimental densities, and LJ parameters were updated via gradient descent using the DMFF framework. The Adam\cite{kingma_adam_2017} optimizer was employed with a fixed learning rate of 0.001. The optimized parameters from one batch were used as the initial guess for the next group, and this process was repeated sequentially until all formulations had been processed. This entire cycle was repeated for a minimum of 20 iterations until the total loss function converged, defined as a relative drift of less than 5
The loss function was defined as the weighted mean squared error between simulated and experimental densities:
\begin{equation*}
Loss = \sum_{i=1}^{N} \omega_i \left( \rho_{\mathrm{sim},i} - \rho_{\mathrm{exp},i} \right)^2,
\end{equation*}
where $N$ is the total number of electrolyte formulations, $\rho_{\mathrm{sim},i}$ is the simulated density, and $\rho_{\mathrm{exp},i}$ is the corresponding experimental density, $\omega_i$ is the weighting factor of the $i$th formulation. A default weight of 1 was assigned to most formulations, while weights of 4 or 8 were applied to formulations containing rare components to ensure their balanced representation in the optimization process and prevent parameter bias toward common electrolyte systems.
\end{itemize}

\section{Supplementary Results}
\subsection{System-Size and Time Dependence of Transport Properties}

\begin{figure}[H]
\centering
\vspace{0.5em} 
\begin{overpic}[width=0.45\textwidth]{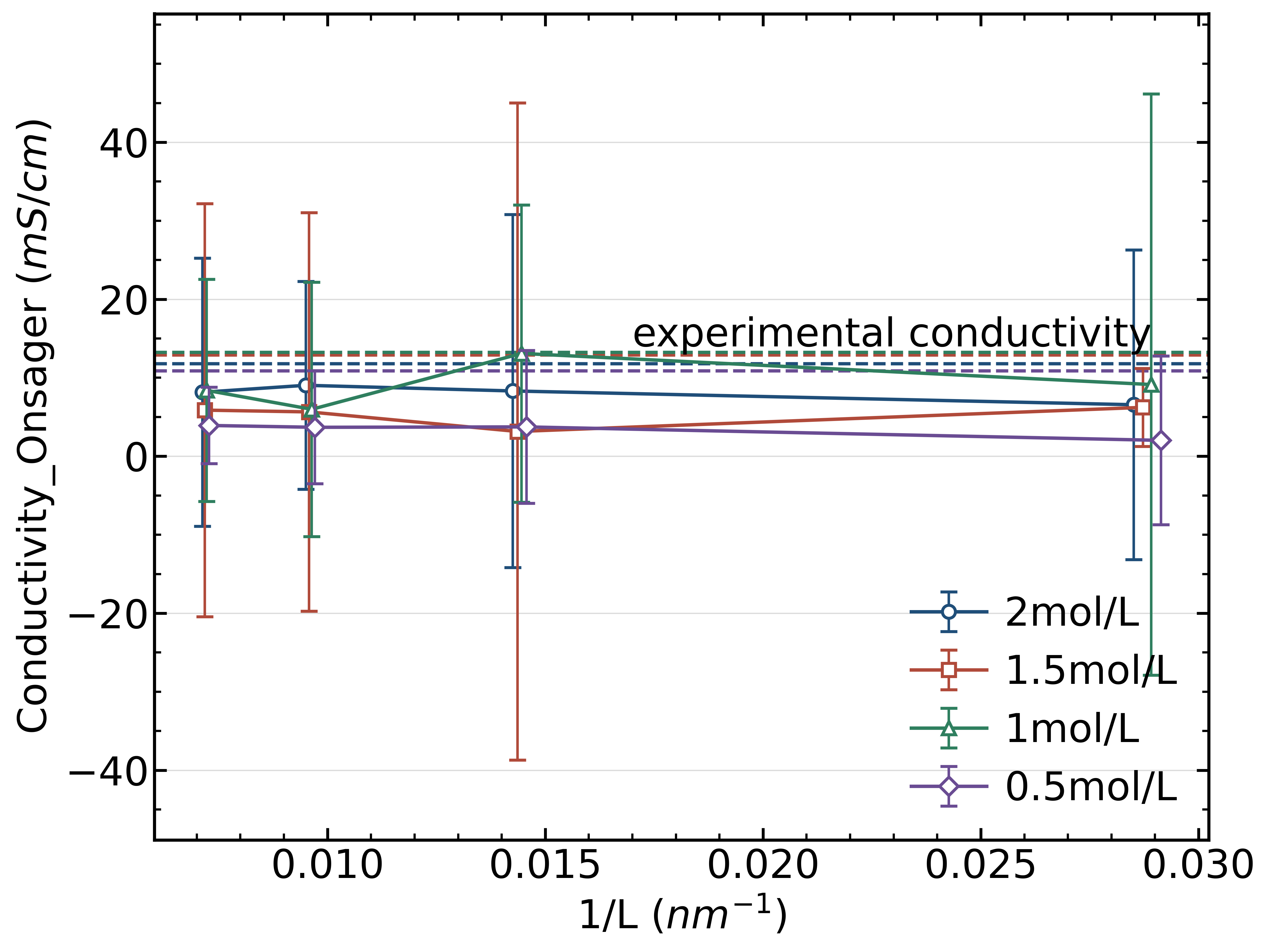}
  \put(-5.0,74){\textbf{(a)}}   
\end{overpic}
\hspace{0.04\textwidth}  
\begin{overpic}[width=0.45\textwidth]{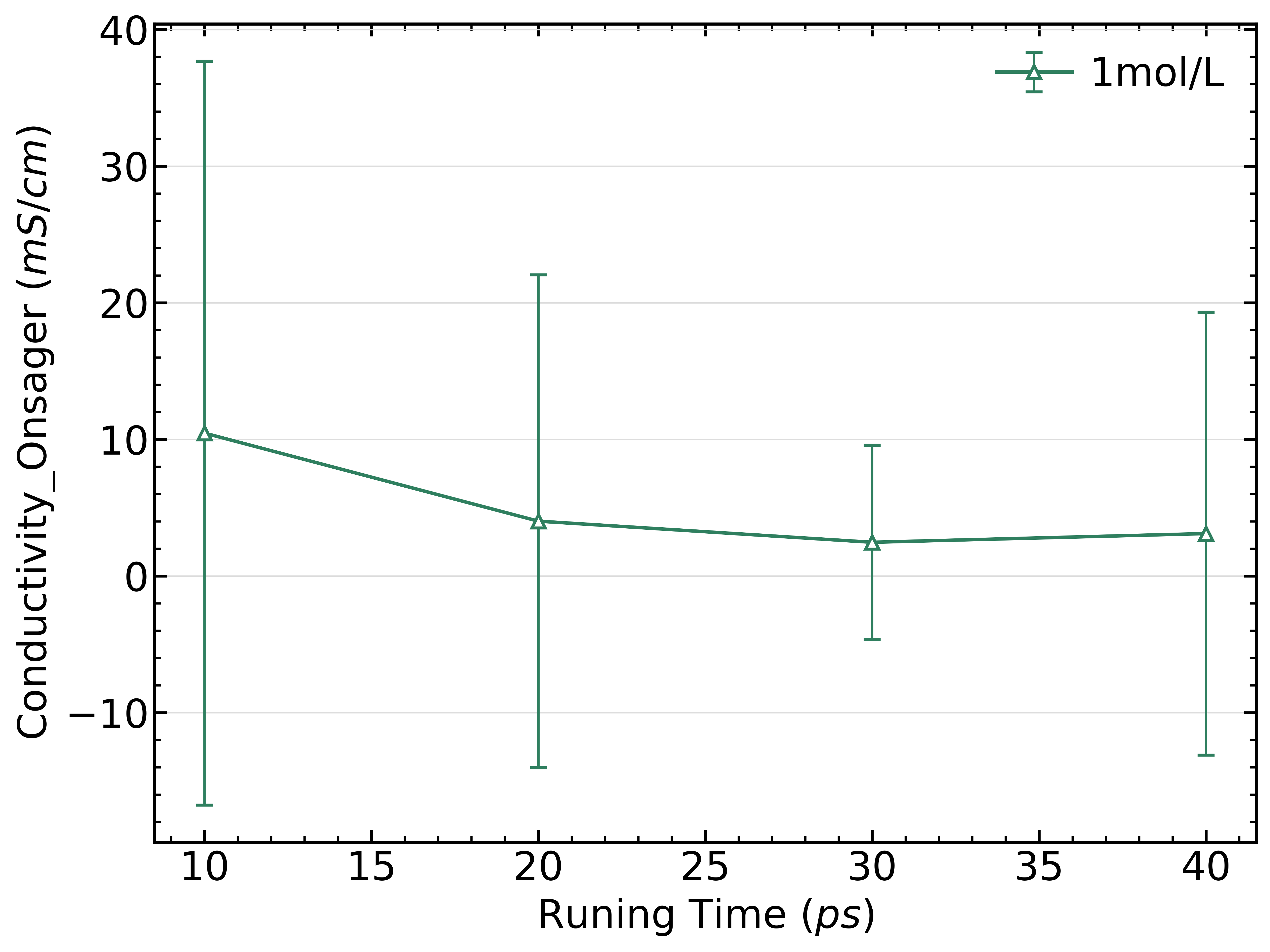}
  \put(-5.0,74){\textbf{(b)}}   
\end{overpic}

\vspace{0.5em} 
\begin{overpic}[width=0.45\textwidth]{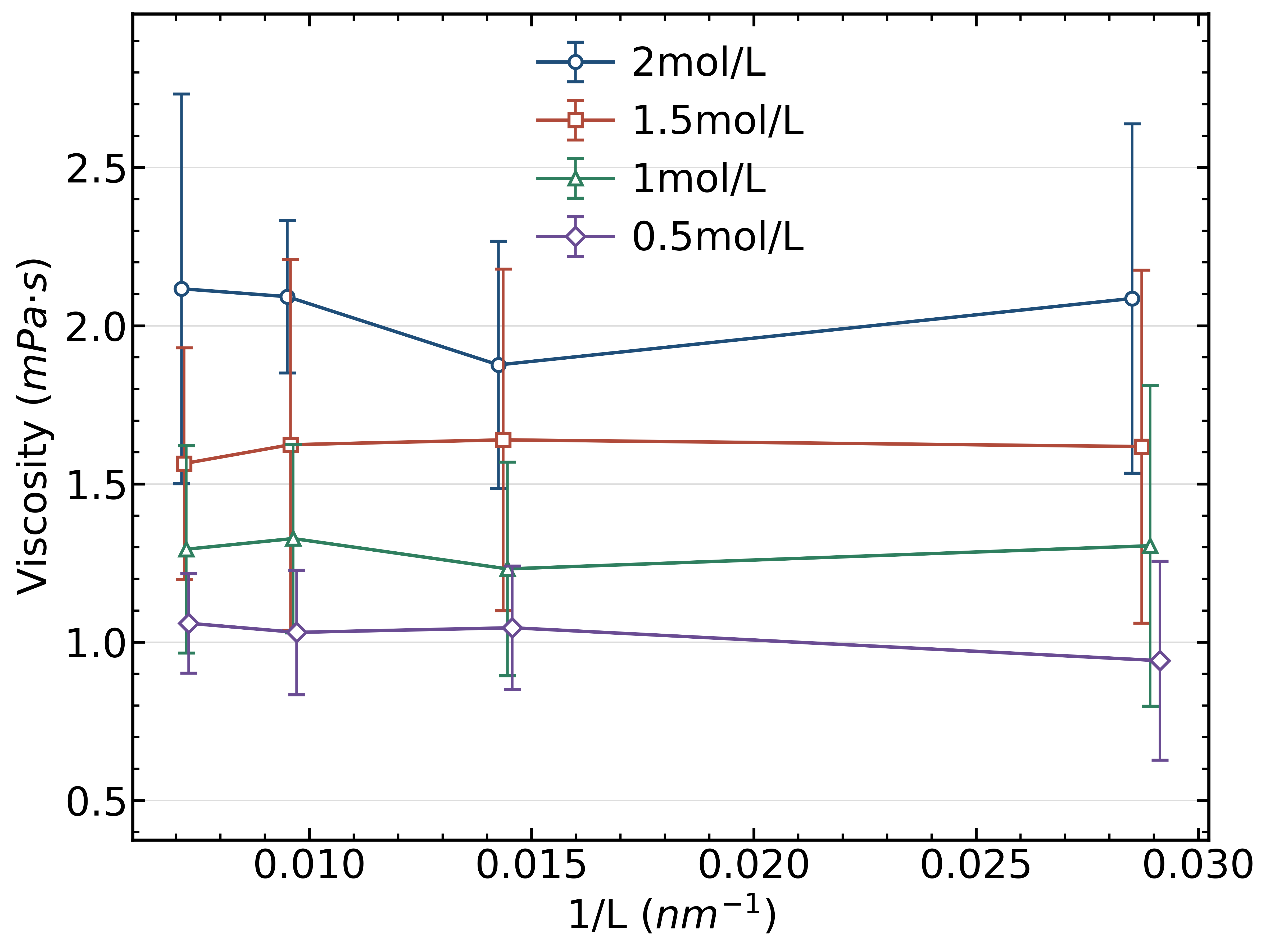}
  \put(-5.0,74){\textbf{(c)}}   
\end{overpic}
\hspace{0.04\textwidth}  
\begin{overpic}[width=0.45\textwidth]{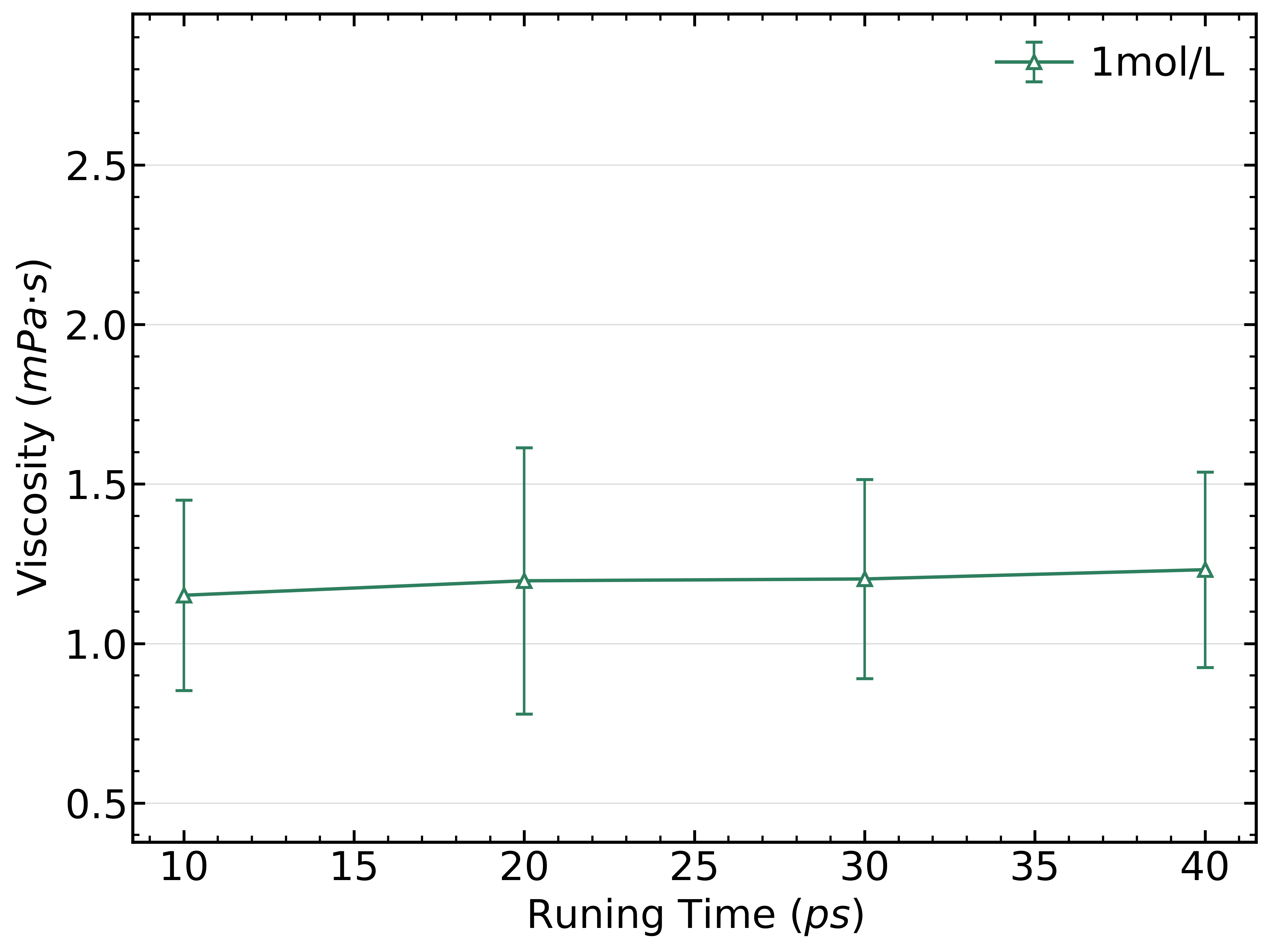}
  \put(-5.0,74){\textbf{(d)}}   
\end{overpic}
\caption{system-size and time dependence of transport properties at different salt concentrations. (a), (c) the convergence of  Onsager conductivity and Green-Kubo viscosity as functions of inverse box length, respectively. Dashed lines in (c) denote experimental conductivity values. Blue, orange, green, and purple symbols correspond to salt concentrations of 2.0, 1.5, 1.0, and 0.5 mol/L, respectively. (b), (d) the time convergence of the corresponding properties.}
\label{fig_s:size_time_convergence}
\end{figure}

Figure \ref{fig_s:size_time_convergence} presents the system-size dependence and temporal convergence of key transport properties (Onsager ionic conductivity and Green-Kubo shear viscosity) across a range of salt concentrations. As shown in Figures \ref{fig_s:size_time_convergence}a and \ref{fig_s:size_time_convergence}c, Onsager conductivity exhibits a clear linear scaling with the inverse box length (1/L), consistent with the finite-size effect of hydrodynamic interactions in periodic systems; in contrast, Green-Kubo viscosity shows no statistically significant dependence on system size within the investigated range, in agreement with previous theoretical and simulation reports. For all salt concentrations, increasing the system size effectively reduces the statistical fluctuation of the calculated transport properties and improves the reliability of the results.
Figures \ref{fig_s:size_time_convergence}b and \ref{fig_s:size_time_convergence}d show the time convergence of Onsager conductivity and viscosity at a representative salt concentration of 1.0 mol/L. Both properties gradually stabilize with extended simulation time: Onsager conductivity converges to a stable value within 30 ns of NVT simulation, and the statistical uncertainty decreases continuously with longer trajectories. These convergence results confirm that the standard simulation protocol (≈10⁵ atoms, 35-40 ns NVT runs) adopted in this work is sufficient to yield reliable and statistically robust transport property predictions.

\subsection{Comparison between GPU and Tianqiong}
    
\begin{figure}[H]
\centering
\begin{overpic}[width=0.45\textwidth]{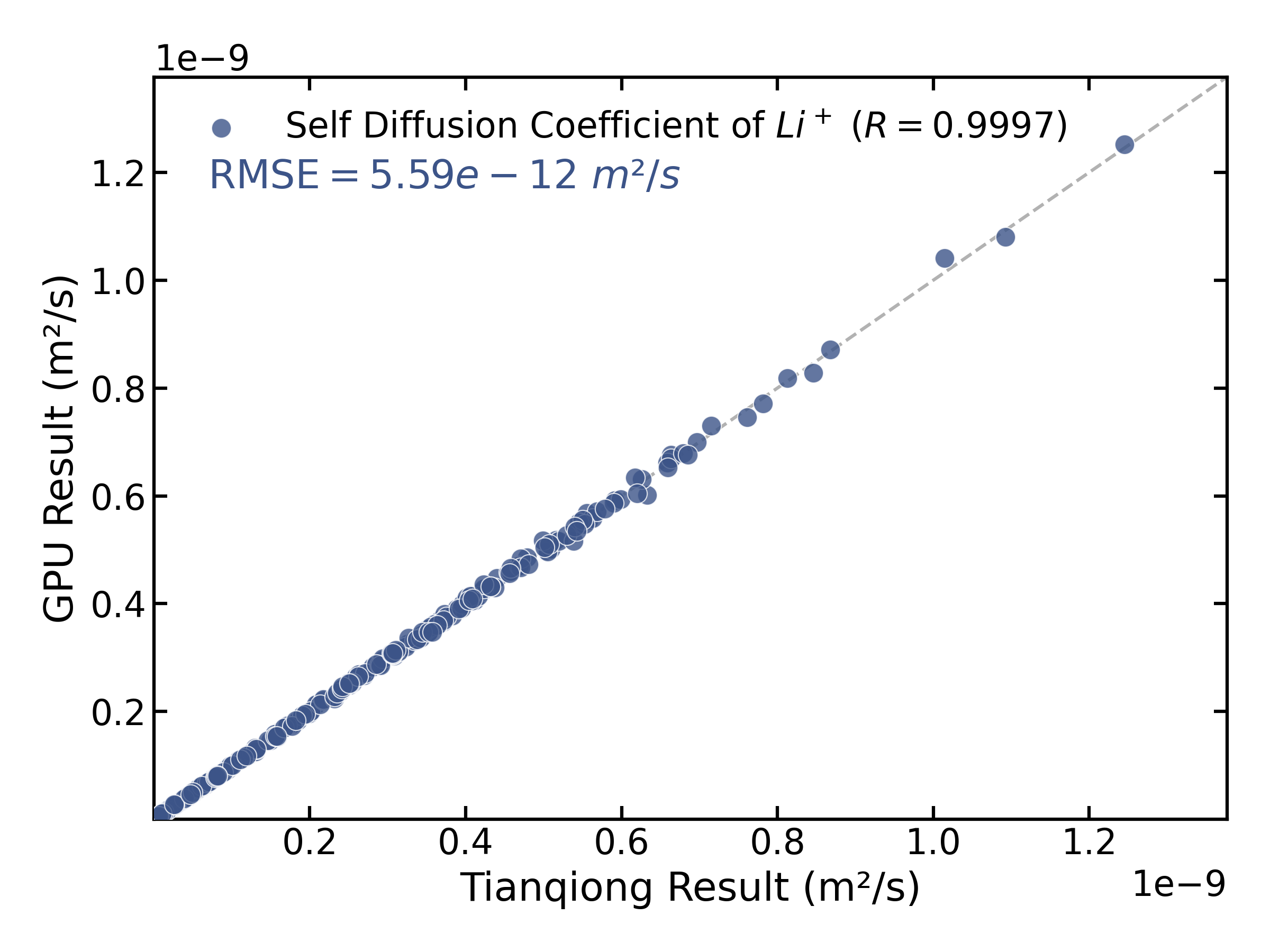}
  \put(0,72){\textbf{(a)}}   
\end{overpic}
\hspace{0.04\textwidth}  
\begin{overpic}[width=0.45\textwidth]{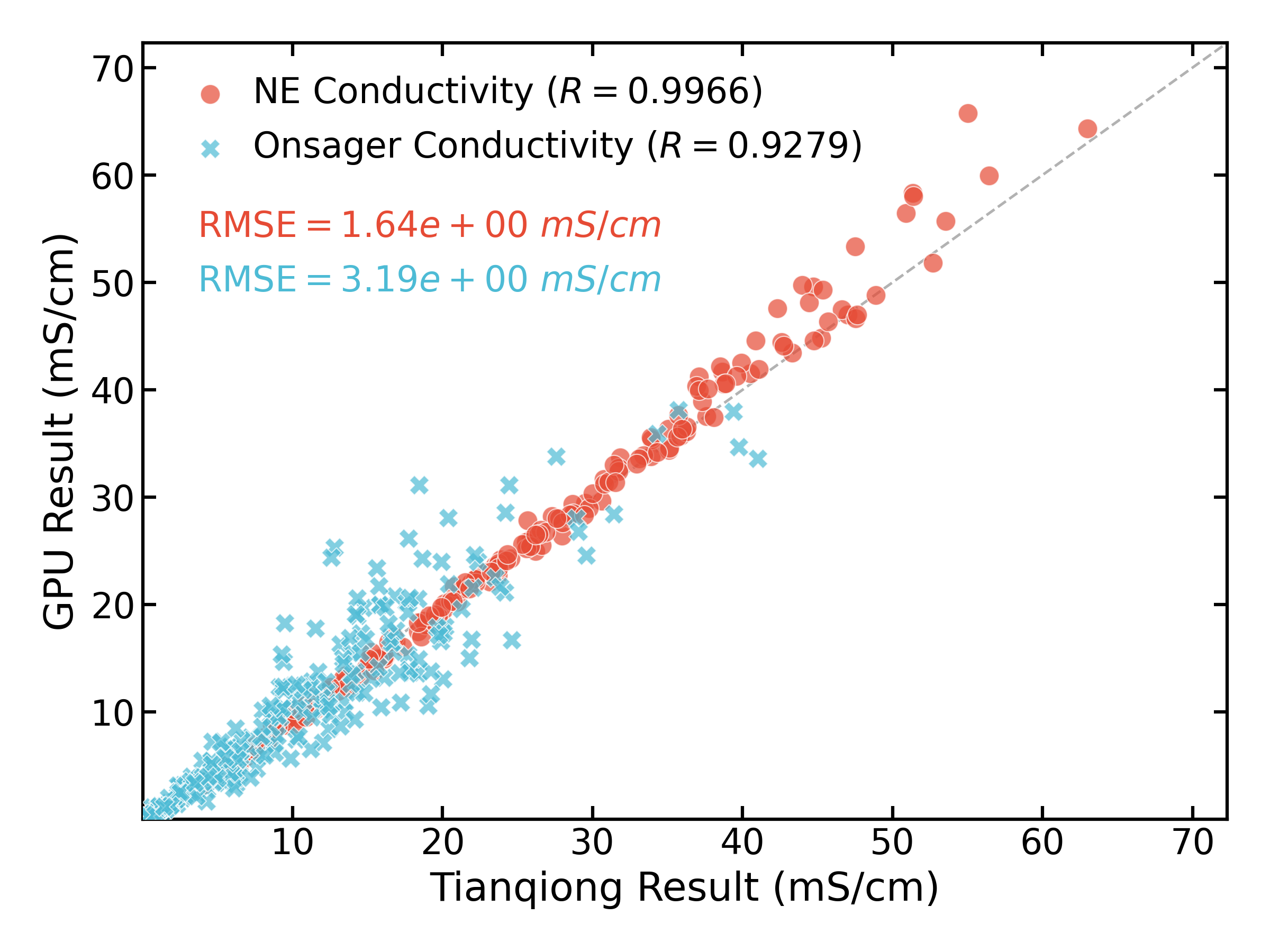}
  \put(1,72){\textbf{(b)}}   
\end{overpic}

\vspace{1em} 
\begin{overpic}[width=0.45\textwidth]{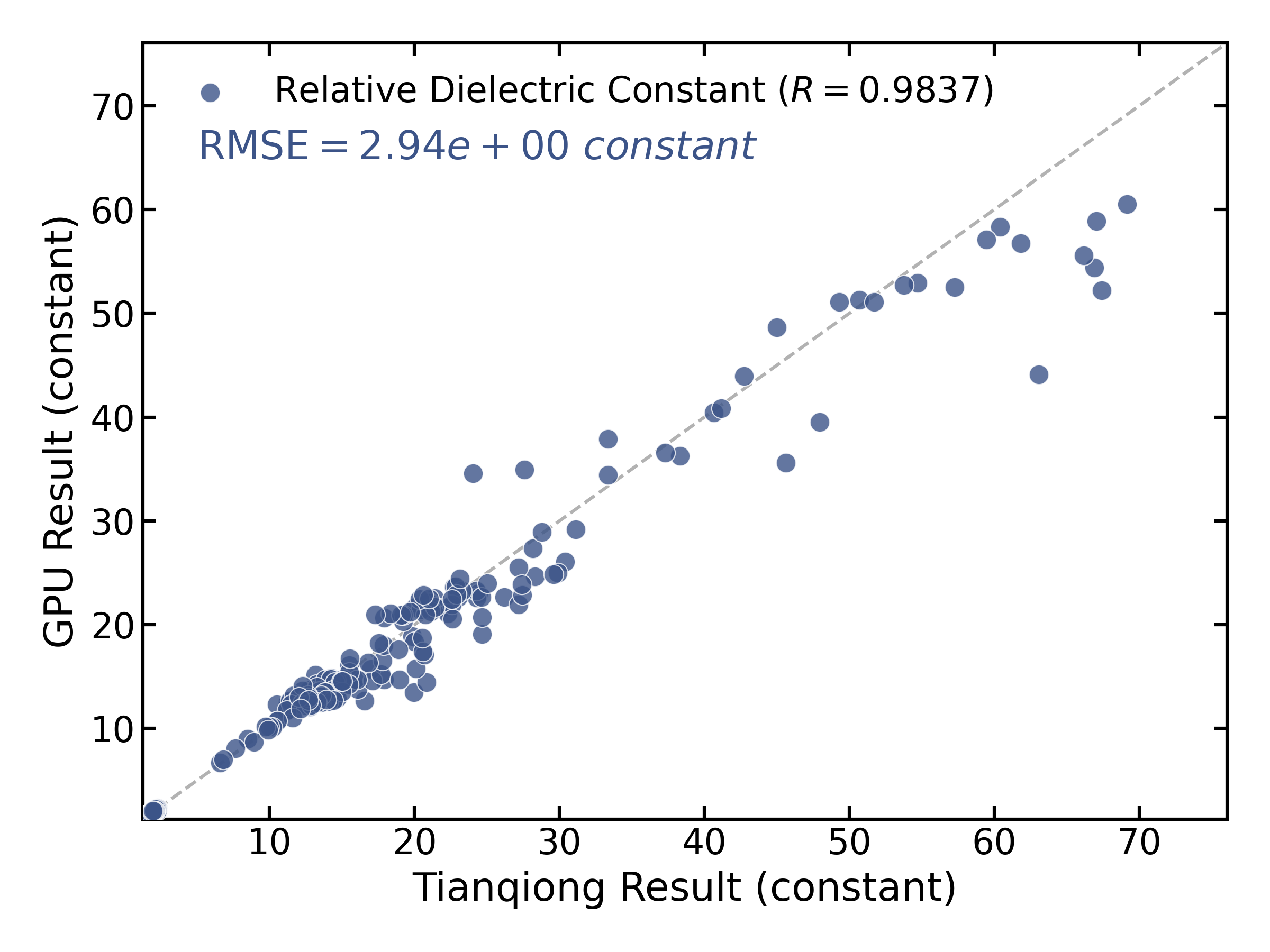}
  \put(1,72){\textbf{(c)}}   
\end{overpic}
\hspace{0.04\textwidth}  
\begin{overpic}[width=0.45\textwidth]{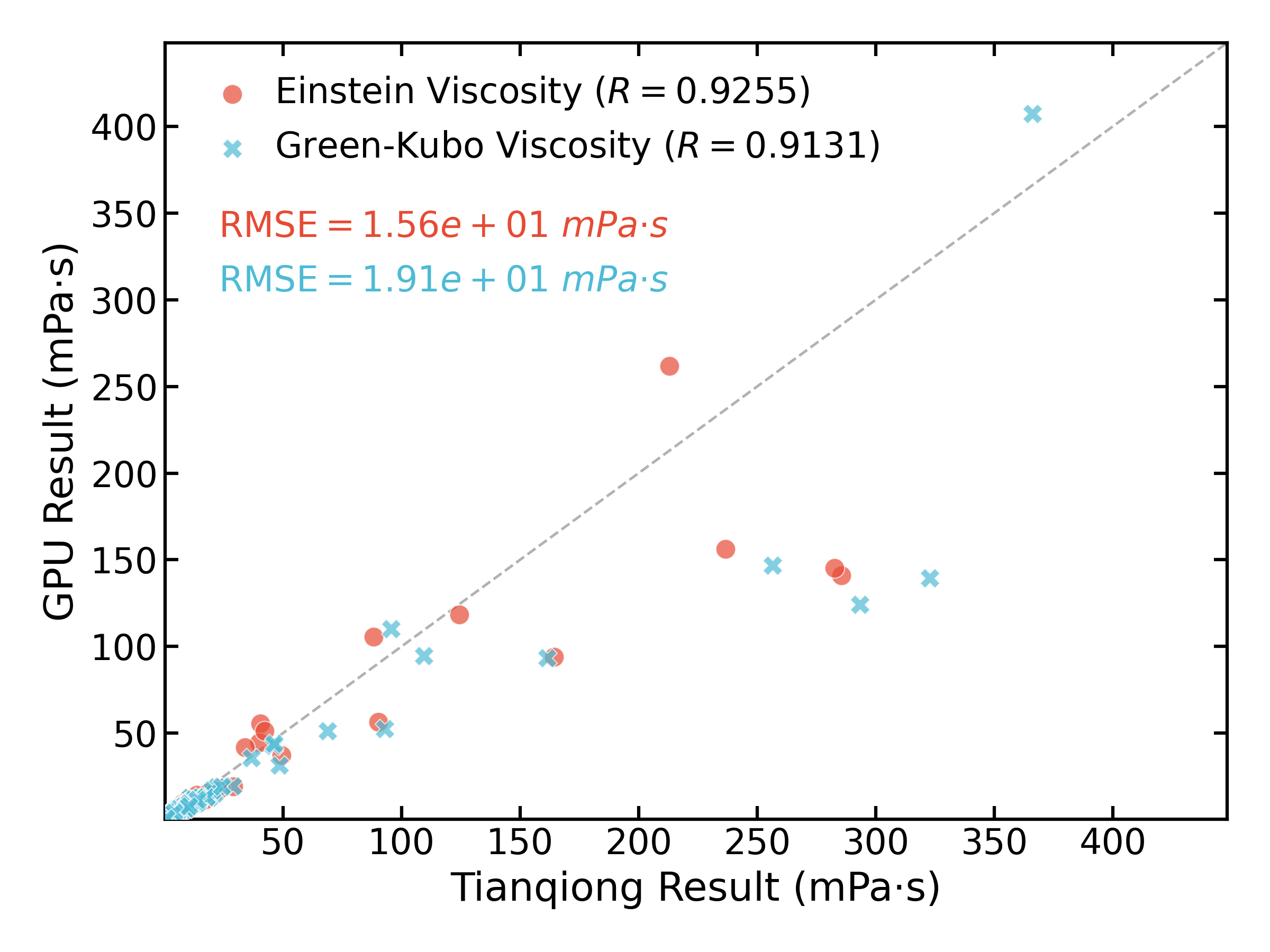}
  \put(1,72){\textbf{(d)}}   
\end{overpic}
\caption{Comparison of prediction properties computed by MD simulations performed by GPU and Tianqiong: (a) self diffusion coefficients of \ce{Li+}; (b) conductivity computed by NE and Onsager method; (c) relative dielectric constant; (d) viscosity computed by Einstein and Green-Kubo method.}
\label{fig_s:GPU_and_Tianqiong_Results_Comparison}
\end{figure}

Figure \ref{fig_s:GPU_and_Tianqiong_Results_Comparison} compares the MD-predicted physicochemical properties obtained from conventional GPU calculations and the Tianqiong platform, including Li⁺ self-diffusion coefficient, Nernst-Einstein (NE) conductivity, Onsager conductivity, relative dielectric constant, Einstein viscosity, and Green-Kubo viscosity. All parity plots show excellent linear agreement between the two platforms, with Pearson correlation coefficients (R) all above 0.91 for all properties. Specifically, the \ce{Li+} self-diffusion coefficient achieves an R value of 0.9998 with an extremely low RMSE, confirming nearly identical calculation results at the atomic trajectory level.

\subsection{t-SNE Visualization of Two Salt Formulation Databases}
\begin{figure}[H]
\begin{overpic}[width=0.9\textwidth]{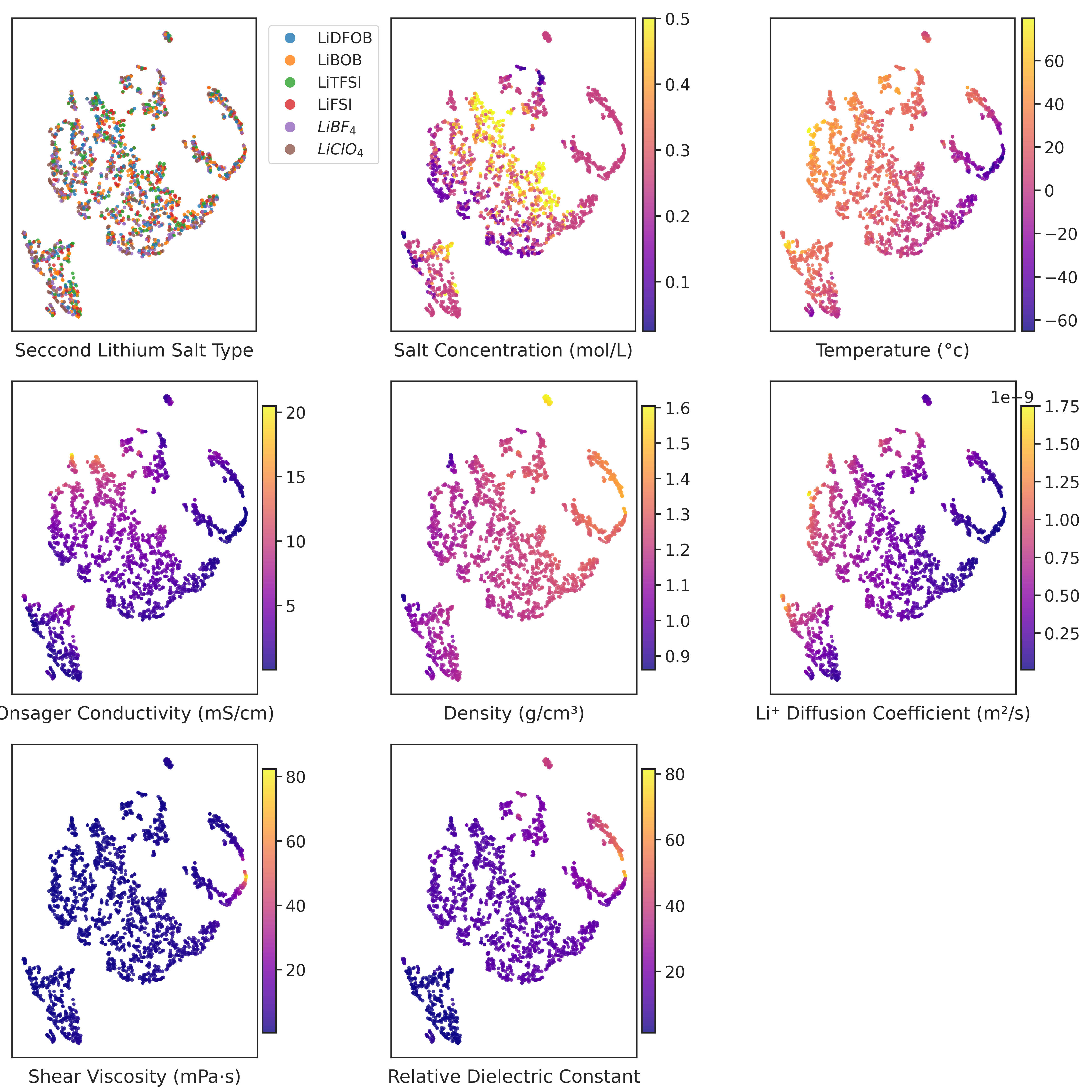}
    \put(1, 95){\textbf{(a)}}   
    \put(36, 95){\textbf{(b)}}  
    \put(71, 95){\textbf{(c)}}  
    
    \put(1, 62){\textbf{(d)}}   
    \put(36, 62){\textbf{(e)}}  
    \put(71, 62){\textbf{(f)}}  
    
    \put(1, 29){\textbf{(g)}}   
    \put(36, 29){\textbf{(h)}}  

\end{overpic}
  \caption{t-SNE 2D embedding of the dual-salt electrolyte property database. All embeddings are calculated from five core physicochemical properties: density, Onsager ionic conductivity, \ce{Li+} self-diffusion coefficient, shear viscosity, and relative dielectric constant. The panels are colored by (a) salt type (highlighting \ce{LiDFOB}, \ce{LiBOB}, \ce{LiTFSI}, \ce{LiFSI}, \ce{LiBF4}, \ce{LiClO4}), (b) same salt concentration for each salt, (c) temperature, (d) MD-predicted Onsager ionic conductivity, (e) density, (f) \ce{Li+} self-diffusion coefficient, (g) shear viscosity, and (h) relative dielectric constant. Green-Kubo viscosity calculations frequently fail to reach full convergence for systems at low temperatures and high salt concentrations. As the t-SNE embedding requires complete data for all five properties per sample, formulations with non-convergent viscosity trajectories were excluded entirely, yielding a final set of 2180 valid duale-salt electrolyte formulations out of 2202 total candidates.} 
  \label{fig_s:t-SNE projection}
\end{figure}

Figure \ref{fig_s:t-SNE projection} shows the two-dimensional t-SNE embedding of the dual-salt electrolyte property database, constructed from five core physicochemical properties (density, Onsager ionic conductivity, Li⁺ self-diffusion coefficient, shear viscosity, and relative dielectric constant). As illustrated in Figure \ref{fig_s:t-SNE projection}a, formulations with different secondary lithium salts exhibit partial regional clustering in the embedding space: sulfonimide-based salts (LiTFSI, LiFSI) tend to locate in regions with moderate conductivity and viscosity, while borate-based salts (LiBOB, LiDFOB) are mainly distributed in high-density and high-viscosity regions, reflecting the influence of anion volume and intermolecular interactions on bulk properties.
Colored by salt concentration (Figure \ref{fig_s:t-SNE projection}b) and temperature (Figure \ref{fig_s:t-SNE projection}c), the embedding reveals continuous gradient distributions of overall electrolyte properties: higher salt concentrations correspond to increased density and viscosity as well as reduced ion diffusivity, while elevated temperatures lead to enhanced conductivity and decreased viscosity. Panels \ref{fig_s:t-SNE projection}d–\ref{fig_s:t-SNE projection}h further visualize the spatial distribution of individual properties, verifying the internal consistency between different physicochemical descriptors (e.g., the inverse correlation between diffusivity and viscosity). Notably, the dual-salt property space partially overlaps with the single-salt database while extending to previously uncovered property regions, demonstrating that dual-salt formulation is an effective strategy to tune electrolyte performance.

\subsection{Distribution of Prediction Properties}

\begin{figure}[H]
\centering
\begin{overpic}[width=0.42\textwidth]{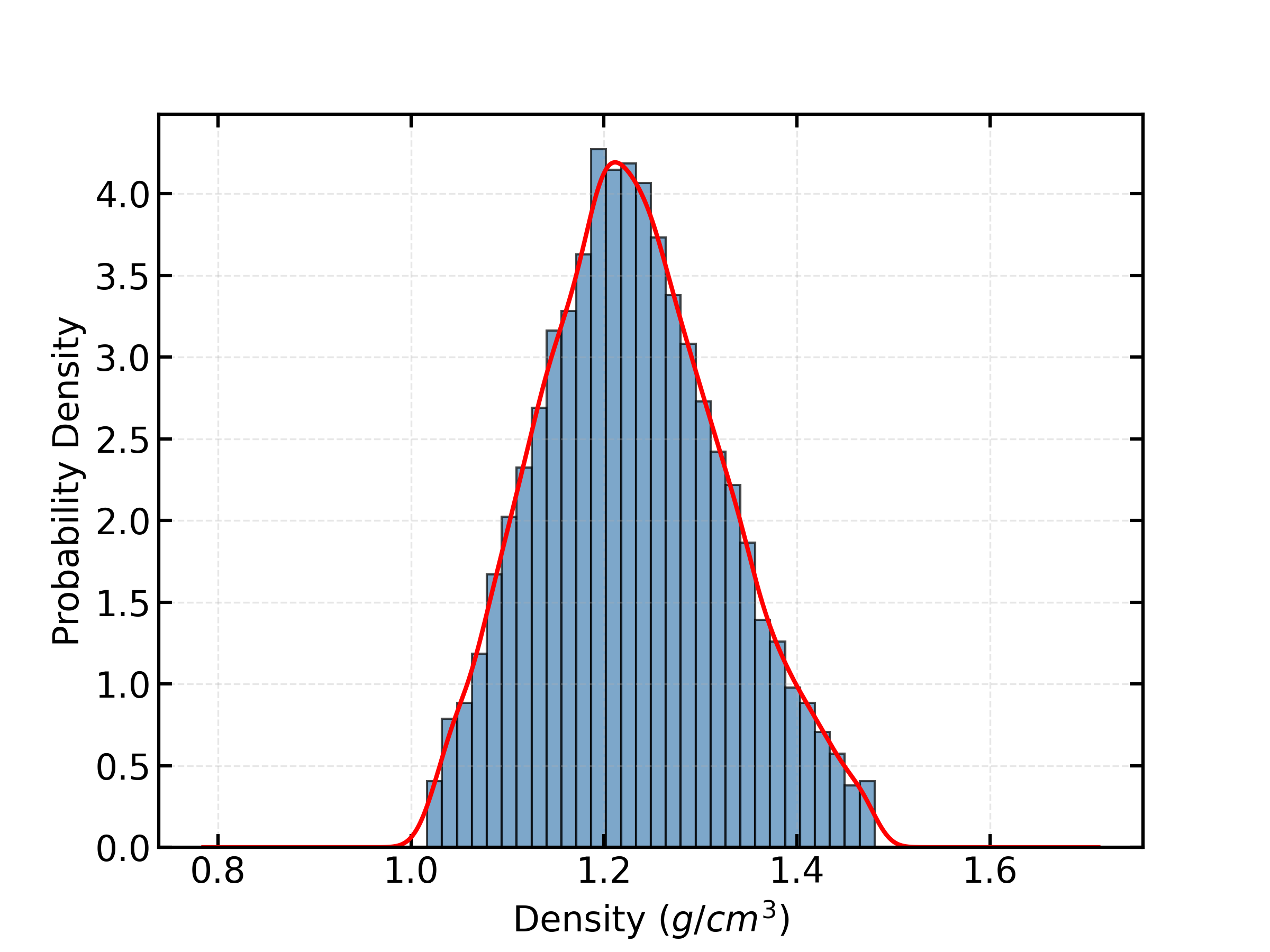}
  \put(-1,65){\textbf{(a)}}   
\end{overpic}
\hspace{0.04\textwidth}  
\begin{overpic}[width=0.42\textwidth]{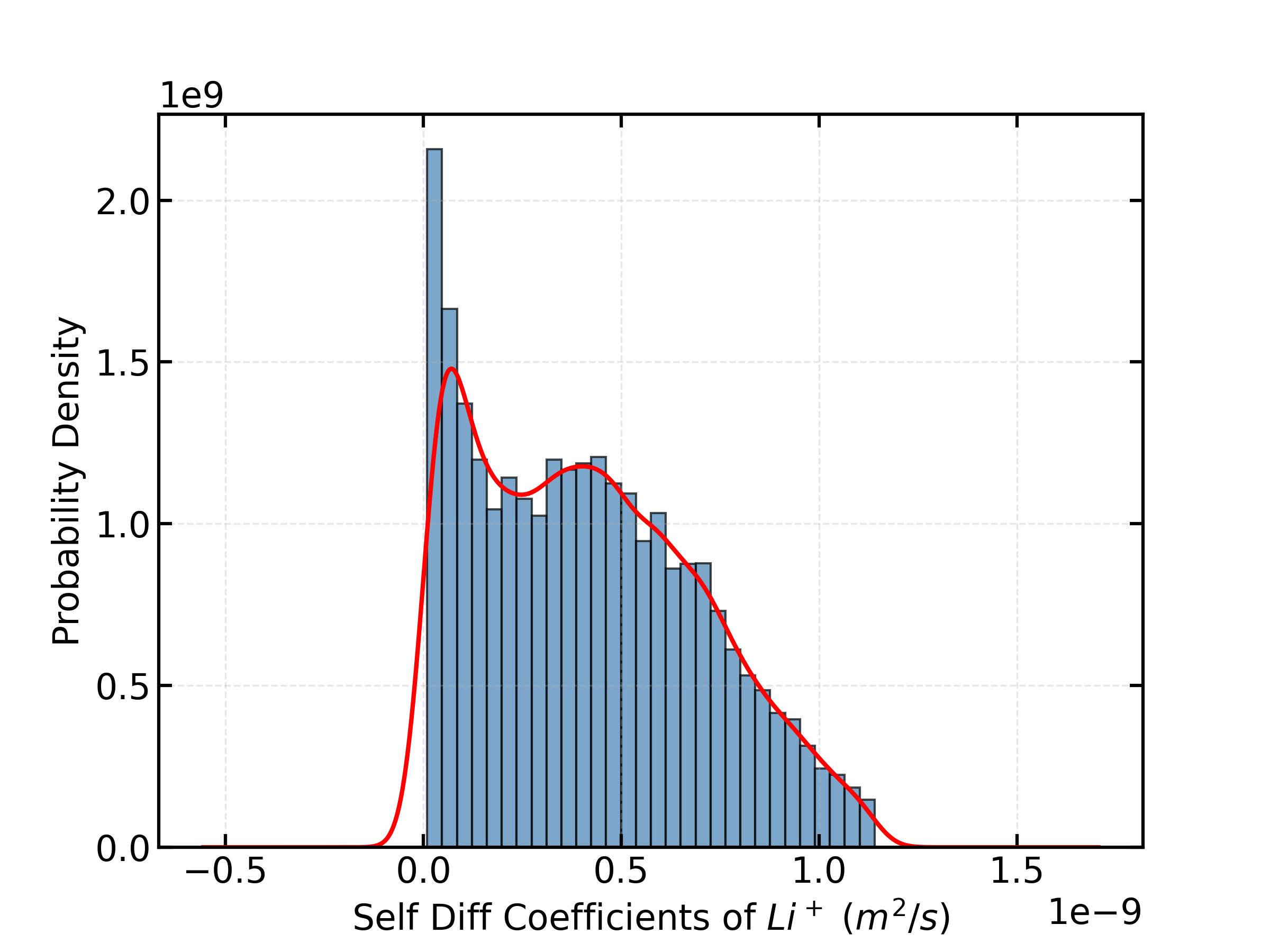}
  \put(-1,65){\textbf{(b)}}   
\end{overpic}

\begin{overpic}[width=0.42\textwidth]{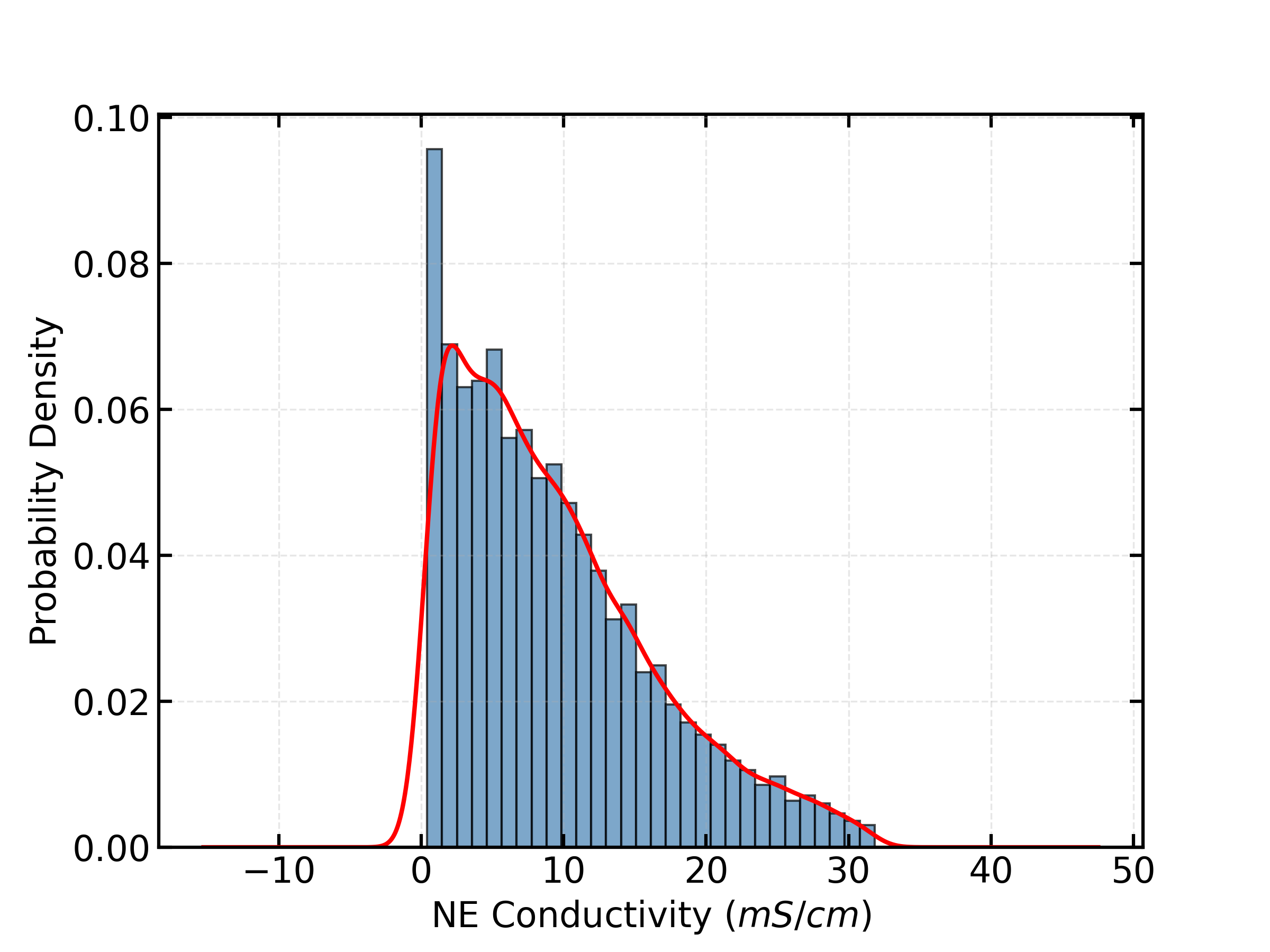}
  \put(-1,65){\textbf{(c)}}   
\end{overpic}
\hspace{0.04\textwidth}  
\begin{overpic}[width=0.42\textwidth]{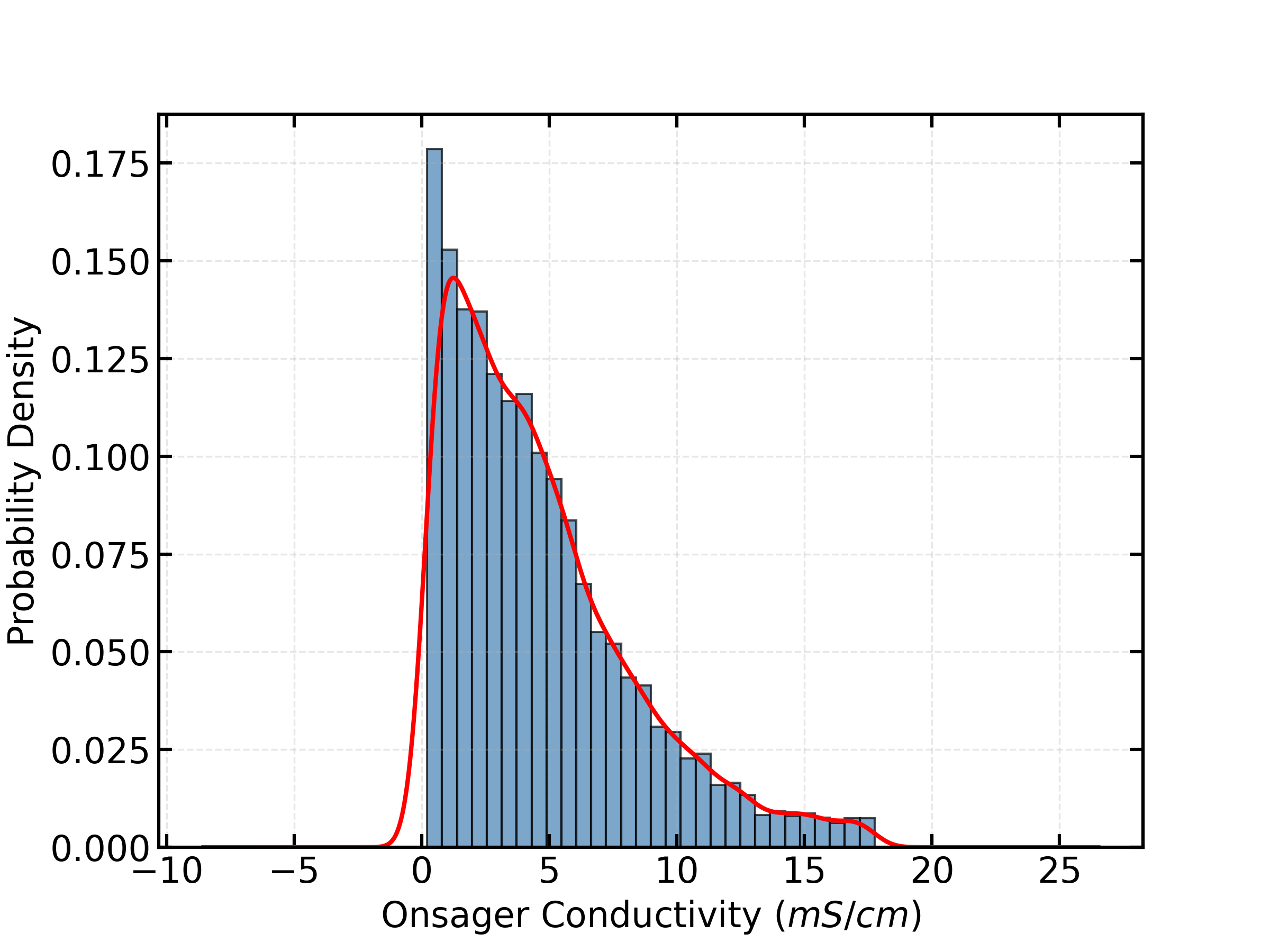}
  \put(-1,65){\textbf{(d)}}   
\end{overpic}

\begin{overpic}[width=0.42\textwidth]{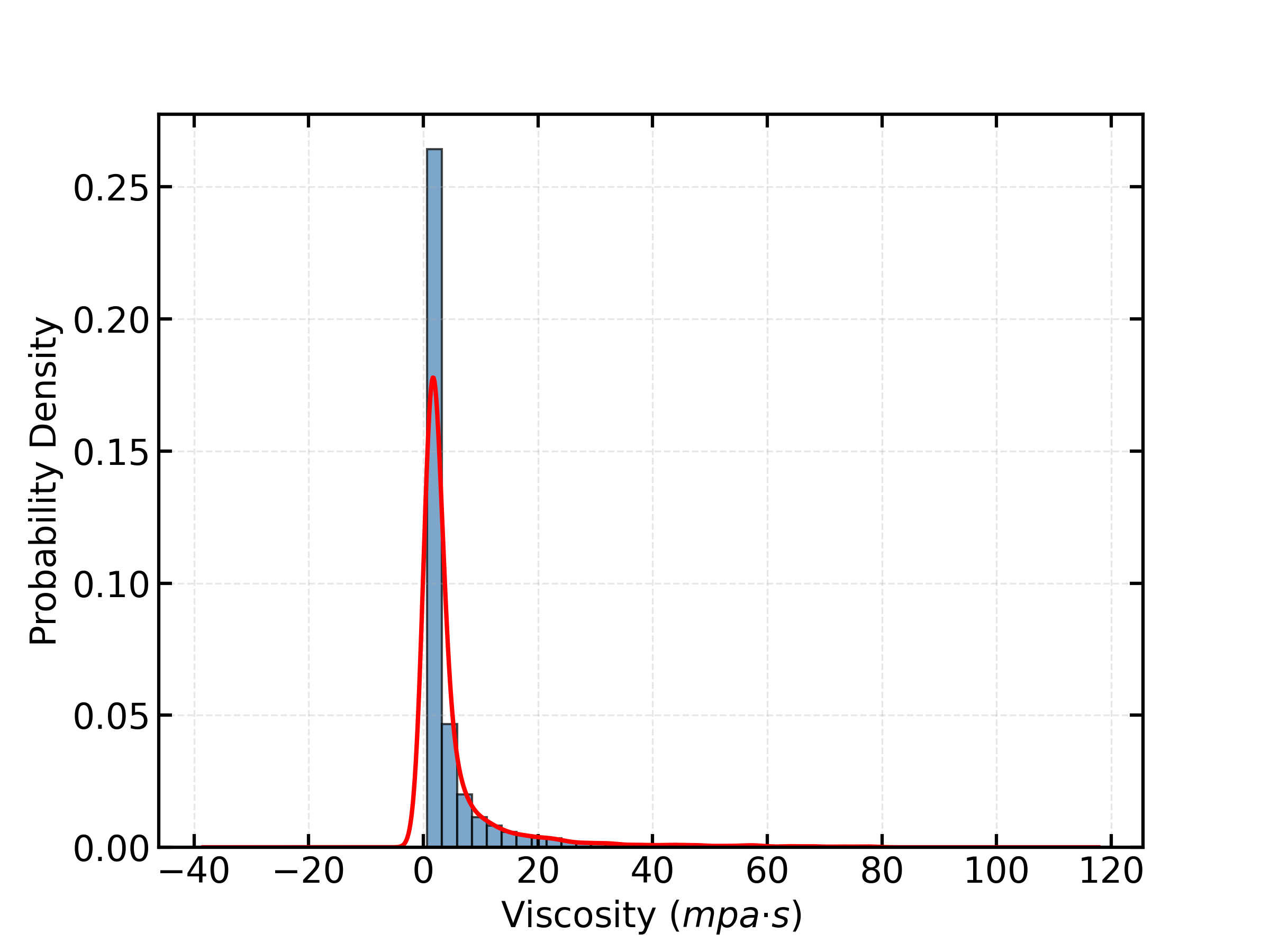}
  \put(-1,65){\textbf{(e)}}   
\end{overpic}
\hspace{0.04\textwidth}  
\begin{overpic}[width=0.42\textwidth]{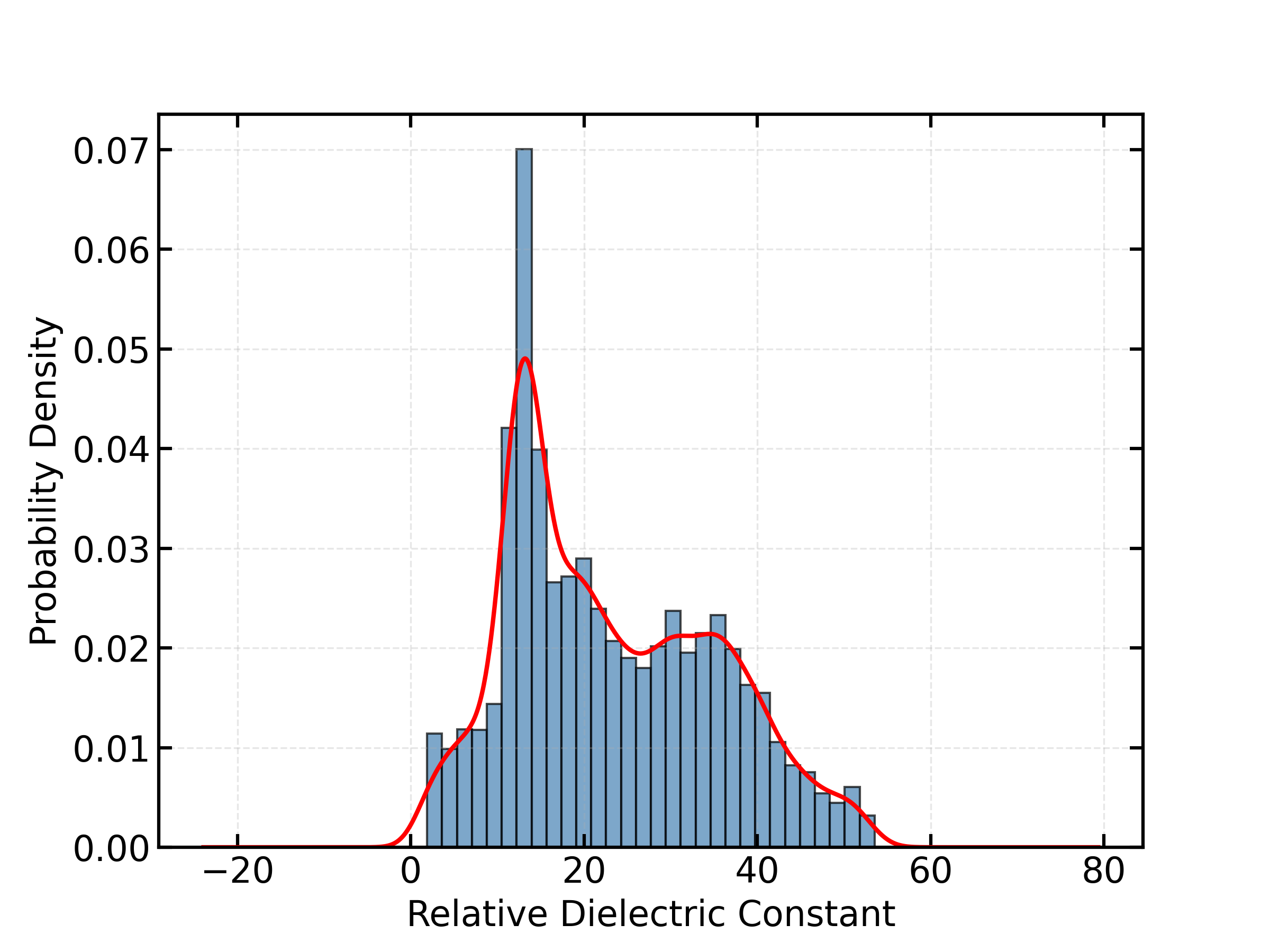}
  \put(-1,65){\textbf{(f)}}   
\end{overpic}
\caption{Distributions of predicted properties for $10^4$ electrolyte formulations: (a) density; (b) Li$^+$ self-diffusion coefficient; (c, d) ionic conductivity computed by the Nernst--Einstein (NE) method and the Onsager method, respectively; (e) shear viscosity computed by the Green--Kubo method; (f) relative dielectric constant. To clearly visualize the main distribution profiles and avoid visual compression caused by long tails from extreme conditions (e.g., ultra-low temperature, ultra-low salt concentration), the top and bottom 2.5\% of extreme values are excluded for each property in the plots.}
\label{fig_s:Distribution_of_Prediction_Properties}
\end{figure}

Figure \ref{fig_s:Distribution_of_Prediction_Properties} displays the probability distributions of six key physicochemical properties for all 10,000 electrolyte formulations in the database. 

\subsection{Screened Salts and Solvents}
67 solvents and 15 salts with the highest occurrence screened from EDB-1 database.
\begin{figure}[H]
\centering
\begin{overpic}[height=18cm]{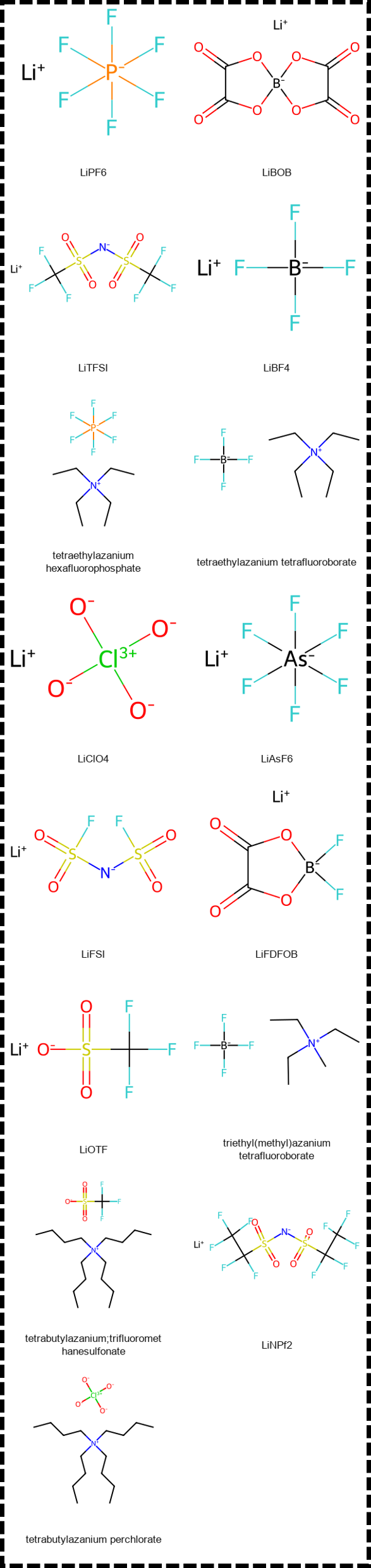}
  \put(-5, 97){\textbf{(a)}}  
\end{overpic}
\hspace{0.06\textwidth}       
\begin{overpic}[height=18cm]{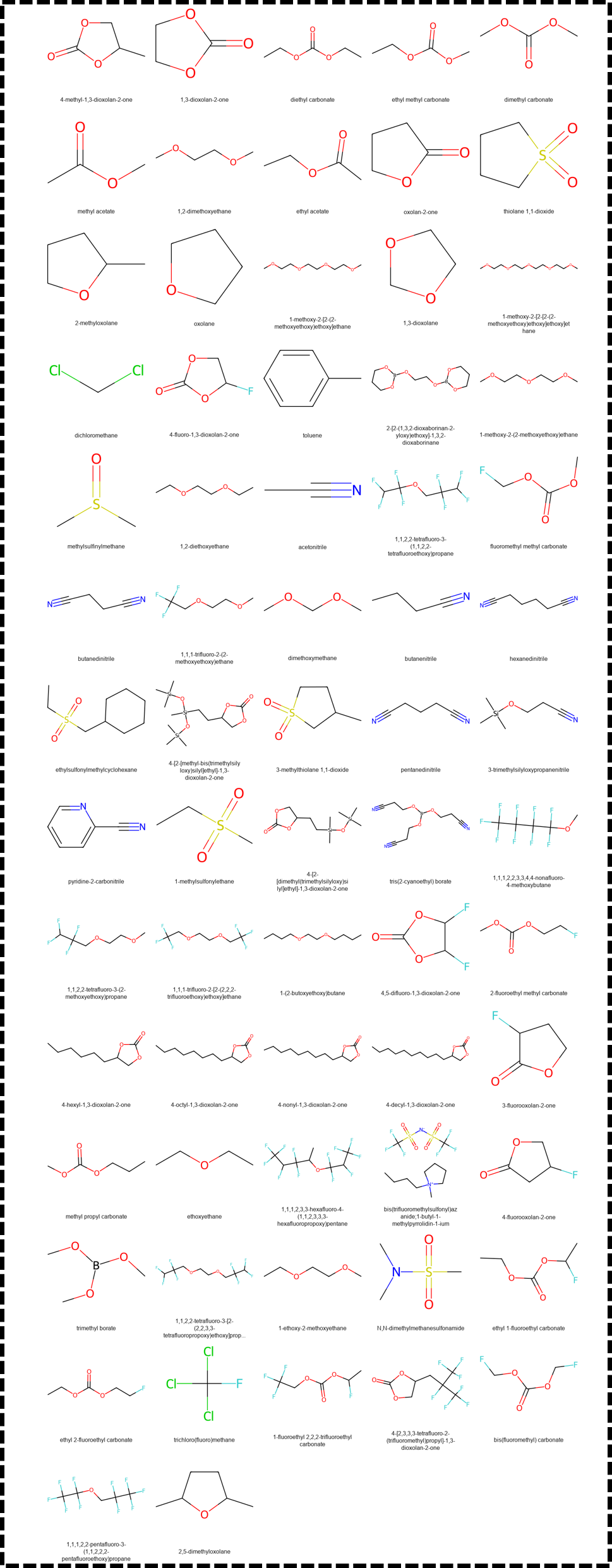}
  \put(-5, 97){\textbf{(b)}}
\end{overpic}

\caption{Screened components from the EDB-1 database. (a) salts; (b) solvents.}
\label{fig:screened_components}
\end{figure}

\bibliography{references}

\end{document}